\documentclass[journal, one column]{IEEEtran}
\usepackage{mathpazo}
\usepackage{times}
\usepackage{amsmath}
\usepackage{amsfonts}
\usepackage{latexsym}
\usepackage{amssymb}
\usepackage{mathabx}
\usepackage{float}
\usepackage{upref}
\usepackage{theorem}
\usepackage{graphicx}
\usepackage{psfrag}
\usepackage{cite}
\usepackage{multirow}
\usepackage{enumerate}
\usepackage{enumitem}

\def\zz{\par\zzz}
\def\zzz#1{%
	\ifx!#1\hfill\else
	\makebox[.75em]{\ifx.#1\else\ifx|#1$|$\else#1\fi\fi}%
	\expandafter\zzz
	\fi}
\usepackage{mdframed}
\usepackage{mathtools}

\usepackage[hidelinks]{hyperref}
\usepackage{graphicx,subfig}
\usepackage[T1]{fontenc}
\usepackage{wrapfig}
\usepackage[font=footnotesize,labelfont=bf]{caption}
\usepackage{soul}
\usepackage{times}

\usepackage{color}
\usepackage{algorithm,algpseudocode}

\makeatletter
\newcommand{\removelatexerror}{\let\@latex@error\@gobble}
\makeatletter
\newcommand{\proofpart}[2]{%
	\par
	\addvspace{\medskipamount}%
	\noindent\emph{Part #1: #2}\par\nobreak
	\addvspace{\smallskipamount}%
	\@afterheading
}
\makeatother

\newcommand{\eps}{\epsilon}
\theoremstyle{plain}
\theorembodyfont{\normalfont\slshape}
\newtheorem{thm}{Theorem$\!$}
\newenvironment{theorem}
{\begin{thm}\hspace*{-1ex}{\bf.}}{\end{thm}}

\newtheorem{clm}[thm]{Claim$\!$}
\newenvironment{claim}{\begin{clm}\hspace*{-1ex}{\bf.}}{\end{clm}}

\newtheorem{lem}[thm]{Lemma$\!$}
\newenvironment{lemma}{\begin{lem}\hspace*{-1ex}{\bf.}}{\end{lem}}

\newtheorem{prop}[thm]{Proposition$\!$}

\newtheorem{cor}[thm]{Corollary$\!$}

\newtheorem{defn}[thm]{Definition$\!$}

\newtheorem{xmpl}[thm]{Example$\!$}
\newenvironment{example}{\begin{xmpl}\hspace*{-1ex}{\bf.}}{\hfill $\Box$ \end{xmpl}}

\newtheorem{cnstr}{Construction$\!$}

\newtheorem{rmk}[thm]{Remark$\!$}
\newenvironment{remark}{\begin{rmk}\hspace*{-1ex}{\bf.}}{\end{rmk}}

\newcounter{enumrom}
\renewcommand{\theenumrom}{(\roman{enumrom})}

\makeatletter
\renewcommand{\@endtheorem}{\endtrivlist}
\makeatother

\makeatletter
\renewcommand{\thefigure}{{\@arabic\c@figure}}
\renewcommand{\fnum@figure}{{\bf Figure\,\thefigure}}
\makeatother

\newcommand{\cA}{\mathcal{A}}

\newcommand{\cD}{\mathcal{D}}

\newcommand{\cG}{\mathcal{G}}

\newcommand{\cO}{\mathcal{O}}
\newcommand{\cP}{\mathcal{P}}
\newcommand{\cQ}{\mathcal{Q}}

\newcommand{\cS}{\mathcal{S}}

\renewcommand{\leq}{\leqslant}

\renewcommand{\geq}{\geqslant}
\renewcommand{\Bbb}{\mathbb}

\newcommand{\Cref}[1]{Co\-ro\-lla\-ry\,\ref{#1}}

\renewcommand{\Bbb}{\mathbb}

\newcommand{\E}{{\Bbb E}}

\outer\def\proclaim #1. #2\par{\medbreak
	\noindent{\bf#1.\enspace}{\sl#2\par}%
	\ifdim\lastskip<\medskipamount \removelastskip\penalty55\medskip\fi}

\ifCLASSINFOpdf
\else
\fi

\begin{document}
%
\title{AC-DC: \underline{A}mplification \underline{C}urve \underline{D}iagnostics for \underline{C}ovid-19 Group Testing}
%
%
%

\author{Ryan~Gabrys$^{\dagger}$,~\IEEEmembership{Member,~IEEE,} Srilakshmi~Pattabiraman$^{\dagger}$,~\IEEEmembership{Student Member,~IEEE,} Vishal~Rana$^{\dagger}$,~\IEEEmembership{Student Member,~IEEE,} Jo\~ao~Ribeiro$^{\dagger}$,~\IEEEmembership{Student Member,~IEEE,} Mahdi~Cheraghchi$^{\ddagger}$,~\IEEEmembership{Member,~IEEE,}, Venkatesan Guruswami$^{\ddagger}$,~\IEEEmembership{Fellow,~IEEE,} and Olgica~Milenkovic$^{\ddagger},~\IEEEmembership{Fellow,~IEEE.}$%

\thanks{R. Gabrys, S. Pattabiraman, V. Rana, and O. Milenkovic are with the Department
of Electrical and Computer Engineering, University of Illinois, Urbana-Champaign, Urbana, IL 61801 USA. e-mail: ryan.gabrys@illinois.edu, \{sp16, vishalr, milenkov\}@illinois.edu.}
\thanks{J. Ribeiro is with the Department of Computing, Imperial College London, London SW7 2AZ, United Kingdom. email: j.lourenco-ribeiro17@imperial.ac.uk.}%
\thanks{M. Cheraghchi is with the Department of Electrical Engineering and Computer Science, the University of Michigan, Ann Arbor,MI 48109
USA. email:mahdich@umich.edu. }%
\thanks{V. Guruswami is with the Department of Computer Science, Carnegie Mellon University, Pittsburgh, PA 15213, USA. email: venkatg@cs.cmu.edu.}%
\thanks{All junior authors$^{\dagger}$ are listed in alphabetical order, and all senior authors$^{\ddagger}$ are listed in alphabetical order. All authors are supported by NSF grant CCF-2107344. Additionally, M. Cheraghchi is supported in part by
an NSF grant CCF-2006455.
}
}

\maketitle

\begin{abstract}
The first part of the paper presents a review of the gold-standard testing protocol for Covid-19, real-time, reverse transcriptase PCR, and its properties and associated measurement data such as amplification curves that can guide the development of appropriate and accurate adaptive group testing protocols. The second part of the paper is concerned with examining various off-the-shelf group testing methods for Covid-19 and identifying their strengths and weaknesses for the application at hand. The third part of the paper contains a collection of new analytical results for adaptive semiquantitative group testing with probabilistic and combinatorial priors, including performance bounds, algorithmic solutions, and noisy testing protocols. The probabilistic setting is of special importance as it is designed to be simple to implement by nonexperts and handle heavy hitters. The worst-case paradigm extends and improves upon prior work on semiquantitative group testing with and without specialized PCR noise models.
\end{abstract}

\begin{IEEEkeywords}
Adaptive Group Testing, RT-PCR. 
\end{IEEEkeywords}

\IEEEpeerreviewmaketitle

\section{Introduction}\label{intro}

\IEEEPARstart{I}{n} less than ten months since the first case reported in the Hubei province of China, Covid-19 has rapidly spread across all continents except Antarctica~\cite{InMap}. The disease has caused more deaths than Ebola, SARS, and the seasonal Flu combined (reaching $1,000,000$ mortalities in September 2020), disrupted the global economy to an extent not seen since the Great Depression and altered the lives of hundreds of millions of people across the globe~\cite{Comp}. 
	
	Many analyses associated with the Covid-19 pandemic have established that widespread population testing is key to effectively containing outbreaks of this and other infectious diseases. In May 2020, the United States was able to test around $150,000$ people per day (According to the Covid Tracking Project, this number has since increased to $750,000$ in August), while countries that have managed to keep the outbreak under control, such as Germany and South Korea, have performed millions of tests during the same stage of the spread of the disease. Although there is no general consensus on the exact number of individuals that need to be tested, most experts agree that the reported numbers are highly inadequate and should be at least an order of magnitude higher before the economy can be safely reopened to the pre-pandemic extent~\cite{Testing}. Some universities, such as Yale University, and the University of Illinois, currently have a biweekly test schedule in place for all individuals accessing school property~\cite{sciencemag}. This is believed to be a sufficiently large-scale testing protocol that allows the institution to safely operate.
	
	To address the need for sustainable high-frequency population testing, a number of countries and states proposed and implemented \emph{group testing schemes} in which genetic samples from different individuals are pooled together in a manner that incorporates thresholded real time reverse transcriptase polymerase chain reaction (RT-PCR) fluorescence signals into the testing scheme\footnote{A recent ordinance issued by the governor of Nebraska~\cite{nebraska} recommends using group testing for widespread screening for Covid-19, while group testing methods are employed in part in Israel.}.
	
	\textbf{Group testing (GT)} is a combinatorial screening method introduced by Dorfman~\cite{Dor43} for identifying small groups of soldiers infected by Syphilis. 
	His scheme, known as single-pooling, consists of mixing blood samples from five soldiers at a time, and running one test for each pool. For positive test outcomes, the soldiers involved in the test are examined individually in a second round to determine who has the disease. For negative outcomes, all subjects involved are declared healthy and removed from future testing schedules. 
Given a relatively small number of infected individuals in a population, this scheme provides a significant reduction in the number of tests required when compared to individual testing~\cite{AJS19}. The scheme proved ineffective for its original task as blood sample pooling dilutes the resulting sample to a point below the sensitivity of the tests used.
	
	A number of recent reports suggest using Dorfman's or other mostly off-the-shelf GT schemes for Covid-19 testing~\cite{ABM+20SinglePooling,BK20,Gol20,Gos20,HT20,NHL20,Tau20NAGT,SLS+20NAGT,YAS+20SinglePooling,ZRB20}. 	Most of the proposed schemes do not incorporate relevant biological priors or exploit the highly specific measurement and noise properties of the RT-PCR method in their testing schemes. {We argue this is a significant detriment}, as in order to properly execute the effort and avoid dangerous failures, testing schemes {should} to be guided both by mathematical considerations as well as social, clinical, and genomic side information\footnote{As the disease affects people from different age groups, ethnicities and regions in a highly disparate manner~\cite{nyt}; it has also been reported that mortality rates across different populations can deviate by as much as two orders of magnitude~\cite{demo}; furthermore, recent studies also suggest that women exhibit significantly milder symptoms than men as they have more responsive immune systems. The World Health Organization (WHO) has also repeatedly issued testing guidelines that suggest ``suspect influenza should be tested with consideration for geographical, gender and age representativeness'' in order to contain the spread of the disease in real-time.}. This suggests designing Covid-19 group testing schemes that carefully address the following issues:
	\begin{enumerate}		
		\item \textbf{Selection of adequate primers.} As stated in the CDC SARS-CoV-2 testing guidelines~\cite{CDC20}, only two primers are recommended for use in the USA for RT-PCR reactions, selected from the N open reading frame (ORF) of the SARS-CoV-2 genome. It is often hard to predict which regions will have small mutations and it is currently not known how fast the N regions and other primer regions chosen by various countries mutate and how these mutations affect the PCR protocol. To determine the influence of mutations one first needs to determine which regions will remain mostly unaffected by mutations, determine the melting temperatures of the primers~\cite{booth2010efficiency} and their binding affinity to the mutated reference regions. For this purpose, the recent work~\cite{rana2020fast} may be used to guide the primer selection process, while actually recorded mutated N primer regions may be used to estimate the failure rates of the individual PCR tests or model the errors in group PCR protocols due to mutations. These issues will be examined in Sections~\ref{bg} and V.  	
		\item \textbf{Selection of (near-)optimal sample mixing strategies with priors.} If not properly designed, GT schemes may lead to errors that are even more detrimental to the population than no tests at all. Since some individuals, such as health workers, may harbor multiple strains of the virus, and since clinical priors are often readily available (e.g., symptom charts, chest X-ray images) the sampling and mixing approaches should be carefully designed to include the right number and combinations of subjects in order to minimize test errors. This is a complicated issue that will be examined elsewhere.
		\item \textbf{Use of quantitative test outcomes.} 
		RT-PCR experiments provide significantly more information about the viral load or number of infected individuals within the group rather than just a simple binary answer, ``no infected samples'' or ``at least one infected sample''. Except for a handful of works proposing to use quantitative RT-PCR through \emph{Compressed Sensing} (CS)~\cite{YMX20NACS,PBJ20NACS,tapestry}, most reported Covid-19 GT schemes assume binary test outcomes (among them the scheme used in Israel and described in~\cite{Shentaleabc5961}).
Furthermore, practically tested schemes operate in a nonadaptive setting, which is suboptimal and not justified for large-scale testing strategies which use a limited number of PCR machines. Another important issue is that heavy hitters (individuals with very high viral loads) can ``mask'' individuals with smaller loads which makes the use of quantitative information difficult~\cite{Mina2020Rethinking,Arnaout2020SARS}. This masking phenomena, as well as saturation effects present in RT-PCR outputs, can make CS approaches highly susceptible to errors. The focus of this work is to develop schemes that can address these issues in a simple, yet efficient manner. Consequently, our main results pertain to scalable, adaptive and semiquantitative testing methods that can efficiently handle heavy hitters and errors that are specific to RT-PCR systems.
		\item \textbf{Incorporation of social/contact network information.} Due to the highly heterogeneous response of individuals to the infection, diverse infection rates across different geographic and communal regions, the best testing practices have to be guided by infection risk assessments and scores. Such ``network-guided'' schemes are currently known in the GT and CS literature, except for a recent work that uses community information to model correlations and reduce the number of tests required~\cite{nikolopoulos2020community}. Instead, we propose to identify highly infected communities and their neighboring communities rather than all individuals in each infected community. In this case, GT is used jointly with other commonly employed mitigation strategies such as quarantine. ``Heavily infected'' community detection as well as heavy hitter problems arising in Covid-19 are discussed in Section~\ref{twostage}.	
	\end{enumerate}
We argue that the availability of (semi)quantitative test outcomes and the use of adaptive strategies can greatly increase the efficiency and scalability of Covid-19 testing schemes. In that direction, we generalize the Semiquantitative Group Testing (SQGT) strategy~\cite{EM14,EM14b,EM16} to an adaptive setting and devise simple probabilistic adaptive GT methods\footnote{The quantifier ``probabilistic'' may refer to either the individuals being ill according to some probability distribution (usually iid Bernoulli$(p)$, or generalized binomial~\cite{hwang1975generalized} or Poisson~\cite{emad2014poisson}, where the number of infected individuals has a right-truncated Poisson distribution with parameter $\lambda(n) = o(n)$; Dorfman's scheme falls into the first category.) or, the test matrices having entries dictated by a probability distribution (again, usually iid Bernoulli$(q)$). Some papers refer to the former setting as ``group testing with probabilistic priors'' and the latter setting as ``group testing with probabilistic tests.'' Which paradigm we refer to will be clear from the context.} and worst-case adaptive GT schemes that take the specific measurement data noise and quantization properties into account. The SQGT scheme assumes that one cannot tell the exact viral load or number of infected individuals in a pool but only an interval in which the load or number of defectives lies. The setting is a generalization of GT that includes more than two outcomes, or a quantized version of the adder channel/CS approach~\cite{wolf1985born,lindstrom1975determining}. It also represents a generalization of the setting~\cite{d1984generalized} in which only saturation effects are taken into account within the adder model. It is worth pointing out that this is also the first approach that uses actual RT-PCR features and also postulates rigorous models that allow for relating viral loads to actual fluorescence values and analyze the testing schemes rigorously. Other methods that will be reviewed in later sections either completely ignore the RT-PCR measurements or do not properly justify or analyze their proposed models.

\begin{figure}[h]
		\centering 
		\includegraphics[width=13cm]{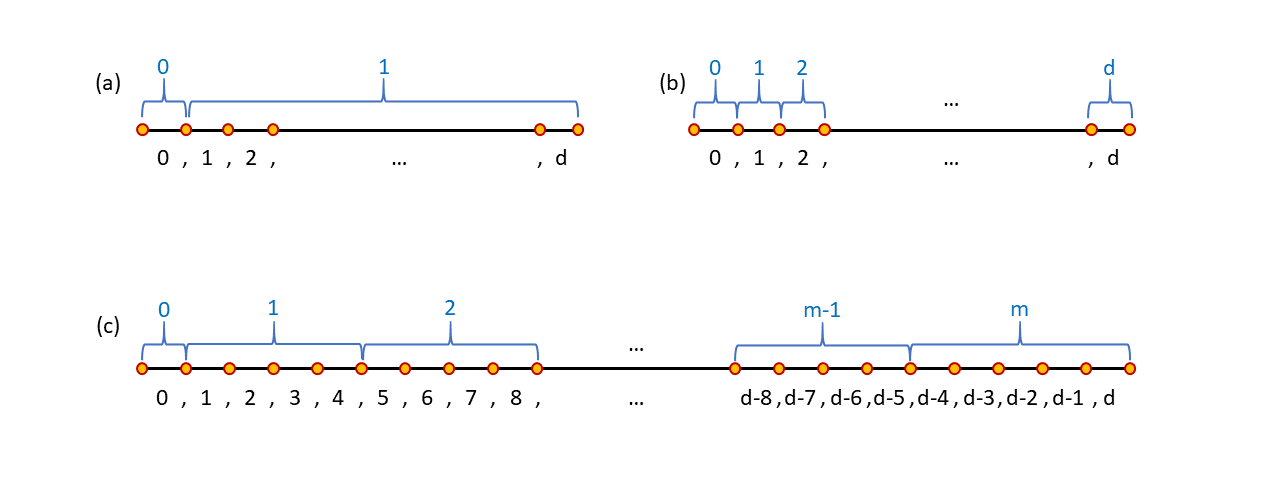}
		\caption{Figure (a) illustrates the \textit{classical GT} framework. Here, $0$ corresponds to a test outcome that is indicative of no infected individual being present, while $1$ corresponds to an outcome indicative of at least one infected individual. Figure (b) represents an \textit{additive test output} model, in which the underlying assumption is that one can tell the exact number of infected individuals in a test. An instance of the \textit{general SQGT} is illustrated in Figure (c). In this case, the test outcome $\lceil \ell/ \tau \rceil=i$ for $i>2$ indicates that $(i-2)\tau \leq \ell \leq  (i-1)\tau$ defectives are present. When the number of defectives detected is $> (m-1)\tau$, the test reports $m$. When $\tau = 1$, (c) represents an adder model with saturation.}\label{fig:comparison}
	\end{figure}

For an illustration of the differences between various testing schemes (group testing, additive tests, additive tests with saturation, and general SGQT), the reader is referred to Figure~\ref{fig:comparison}.

	For \textbf{the worst case setting}, in which we assume a known number of $d$ defectives, but make no assumptions about how they are distributed, and where one can tell if zero defectives were present, or the number of defectives is nonzero and lies in one out of $m$ consecutive intervals of length $\tau$, the number of tests per defective roughly equals
	$$\frac{\log\,(n/d)+\log\log(m+1)}{\log(m+1)}.$$
	The savings in the number of tests as compared to the classical GT setting provided by the increased resolution of the levels is $\log(m+1)$-fold, which even for $7$ levels amounts to three-fold savings. Clearly, one has to be able to properly calibrate the RT-PCR readouts and determine adequate thresholds in order to take full advantage of the scheme. This issue will be addressed in Section~\ref{bg}. 
	
	For \textbf{the probabilistic setting}, where each item is assumed to be independently defective with some probability, our main results include simple-to-implement algorithms for adaptive testing that involve two thresholds and two test stages, and are also capable of handling heavy hitters (i.e., individuals with high viral loads that may mask other individual's presence, provided these individuals are not too common.)
	
	The remainder of the paper is organized as follows. Section~\ref{bg} provides an overview of the PCR, RT-PCR and the Real Time (Quantitative) PCR techniques. The section also addresses key issues that impede the amplification efficiency of the methods currently used for Covid-19 testing and introduces several practical noise models. Section~\ref{bg:gt} describes various GT approaches and assesses their utility for Covid-19 testing. Sections~\ref{twostage} and \ref{sec:worst} describe the main original results. Section~\ref{twostage} describes a probabilistic version of a SQGT model, simplified to account for two rounds of testing and two test thresholds only. This section also introduces test schemes that aim to identify highly virulent individuals, termed \emph{heavy hitters}. Section~\ref{sec:worst} introduces a new worst-case adaptive SQGT technique that is near-optimal and describes a noise model termed the \emph{birth-death chain model}. Section~\ref{ongoing} provides preliminary directions towards designing efficient community-aware testing strategies. 
Section~\ref{modeling} reports the results obtained using the GISAID database~\cite{shu2017gisaid} that explain the influence of mutations in the primer regions on the efficiency of the tests and therefore suggests that noise models previously considered in the literature that only account for errors in the PCR test are inadequate. Section~\ref{sec:conclude} concludes the paper and discusses future work.
	
	\section{Background} \label{bg}
	We start our exposition by describing the real-time reverse transcriptase (RT-PCR) testing mechanism. 
	DNA has a double helix structure and both strands in the helix are composed of periodic sugar and phosphate groups to which one of four different bases is attached, namely A, T, C, and G. A sugar, phosphate, and base are jointly referred to as a nucleotide. As the sugar is asymmetric in terms of the placement of its carbon atoms with respect to the position of binding to the phosphates, the two strands of the DNA have two different directions: One runs from the 3' to 5' carbon end, while the other runs from the 5' to 
3' carbon end. The two strands are held together through the stacking of bases and the hydrogen bonds that exist between them. The pairing rule is dictated by Watson-Crick (WC) complementarity asserting that only (or with overwhelming probability) the bases A and T and G and C bind to each other, respectively. 

\subsection{Reverse transcriptase PCR}
	
\begin{figure}[h]
		\centering
		\vspace{-0.1in}
		\includegraphics[width=12cm]{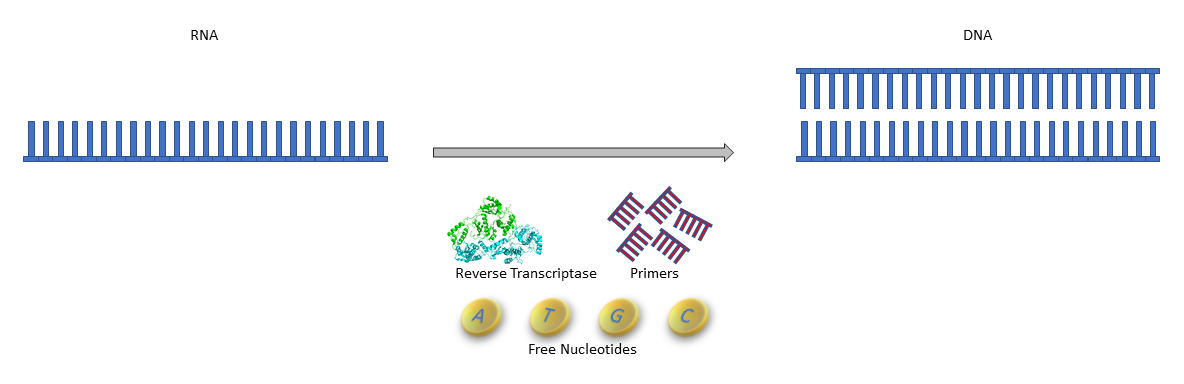}
		\caption{Reverse transcription for converting viral RNA into cDNA.}\label{fig:reverse}
		\vspace{-0.1in}
	\end{figure}

	The Reverse-Transcriptase PCR (RT-PCR) technique is used to identify/amplify RNA strands. Since RNA is single-stranded and hence an unstable molecule, RT-PCR first converts the target RNA into its complementary double-stranded DNA (cDNA, as illustrated in Figure~\ref{fig:reverse}) and then performs amplification using the standard PCR technique. Note that RNA has three of the same building blocks as DNA, namely A, C, and G, but instead of T (encountered in DNA), RNA contains U (Uracil). 

Conversion of RNA into cDNA is accomplished through the use of the reverse transcriptase (RT) enzyme that stitches together ``free'' nucleotides A, T, C, and G together, in the presence of primers that are complementary to a specific part of the target RNA sample (see Figure\footnote{Reverse Transcriptase image is from Wikipedia.}~\ref{fig:reverse} ). Since RNA is inherently single-stranded, the primers have an affinity to attach to the complementary RNA regions, recruit the RT enzyme and thereby initiate synthesis. The process proceeds through two steps: In the first step, the first-strand cDNA is created using the single-stranded RNA as a template. In the second step, the second-strand cDNA is formed by using the first-strand cDNA as template. Consequently, the product cDNA represents an accurate replica of the original RNA content, converted to the DNA alphabet. 

The test results are usually compared to a control as a means to assess the quality of the experiment. RT-PCR is used to detect RNA viruses, i.e., viruses whose genomic content is stored in RNA rather than DNA. SARS-Cov-2, the virus causing Covid-19, is an RNA virus, as is for example the HIV virus that causes AIDS. 
For viral detection, the first step of testing involves isolating genomic RNA by breaking the viral membrane, but this and other processes that lead to actual sample isolation will not be discussed in this short review. 

The Covid-19 detection and amplification process relies on standard RT-PCR methods and RT-PCR and its quantitative version described next. 
	
	\subsection{The Polymerase Chain Reaction (PCR)}
	
\begin{figure}[h]
		\centering
		\includegraphics[width=10cm]{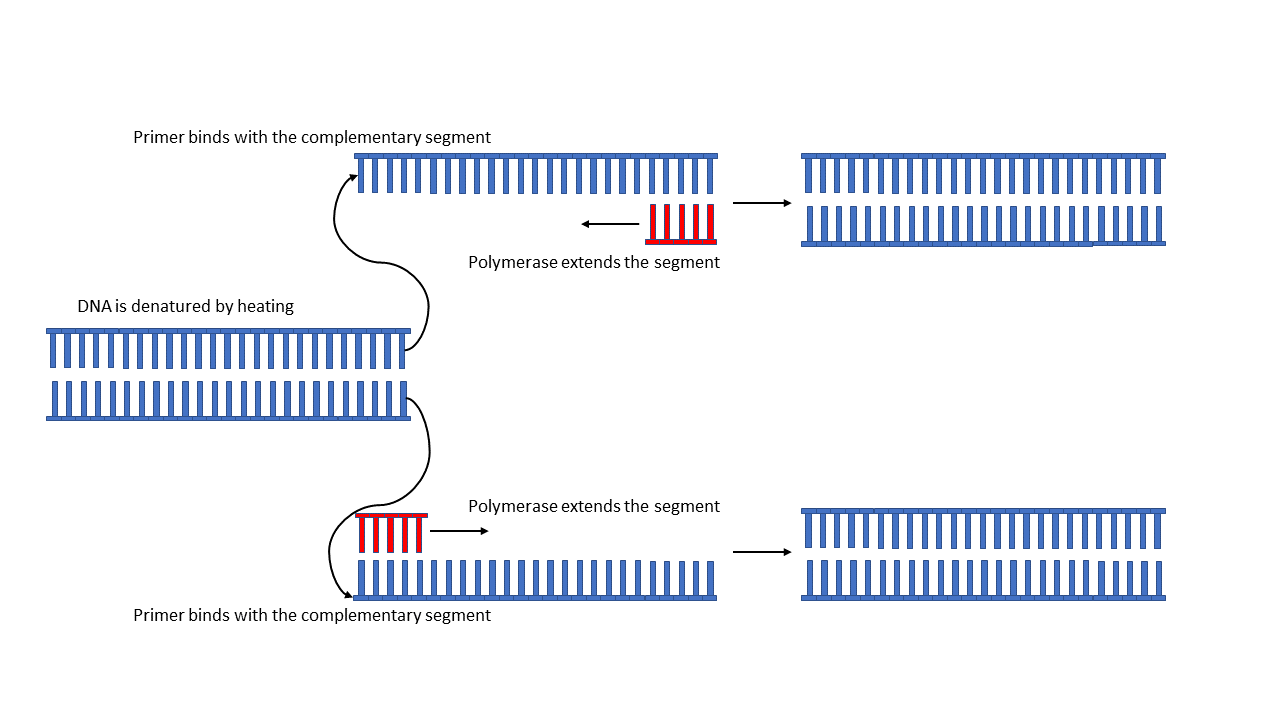}
		\caption{Polymerase Chain Reaction: In any given cycle, the DNA strands in the sample are first denatured into single-strands. The two single-strands are then extended to form complete DNA double-helixes. The primers (short DNA fragments) are attached to the single-stranded DNA such that extension is facilitated in the 3'-5' direction. At the end of each ideally executed cycle, the number of DNA strands in the sample doubles.}\label{fig:pcr1}
	\end{figure}

\begin{figure}[h]
		\centering
		\includegraphics[width=11cm]{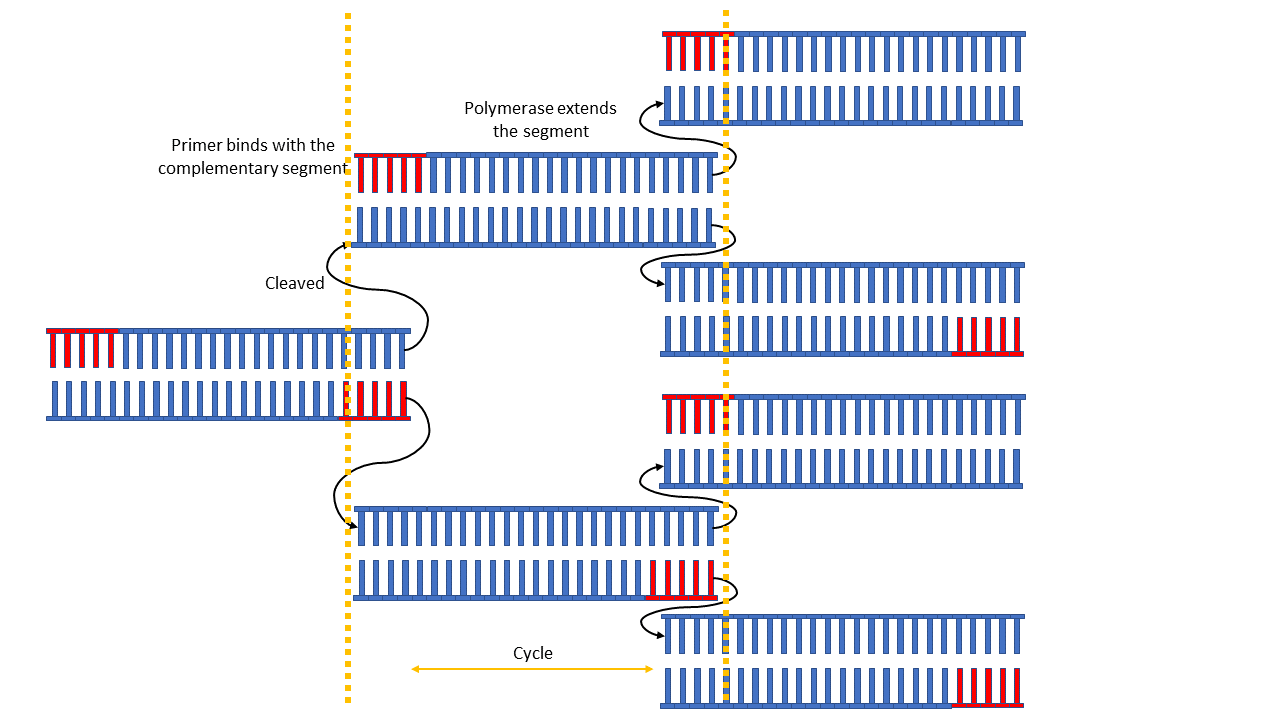}
		\caption{Under ideal conditions, every cycle of the PCR process should double the DNA content. Due to various factors, described in the main text, not every cycle may result in twice as many strands and an averaged efficiency factor $\eta<1$ is used to describe the growth rate of the PCR product.}\label{fig:pcr2}
	\end{figure}

	The {Polymerase Chain Reaction} (PCR) is used to amplify specific segments of the 
	DNA strands in order to enable a downstream analysis of the segments or to detect the presence of specific DNA content. The operating principles of the PCR process are illustrated in Figures~\ref{fig:pcr1} and~\ref{fig:pcr2}. 
	A Thermal Cycler uses the target DNA, specific primers (short DNA segments that initiate the replication process by allowing the polymerase to bind to the DNA), the taq polymerase (which actually performs DNA replication after the primers get attached), and free A (Adenine),T (Thymine), C (Cytosine) and G (Guanine) nucleotides needed to amplify the segment of interest through \emph{repeated cycles} that involve the following steps: \emph{DNA denaturation, annealing, and extension.}
	\begin{enumerate}
		\item DNA Denaturation: The DNA sample to be amplified or detected is first heated to $96^{\degree}$C. At this temperature, hydrogen bonds between the bases across the two strands break, producing two complementary single-stranded DNA fragments.
		\item Annealing (Hybridization): The sample is subsequently cooled to $55^{\degree}$C. This allows the primers to bind to their WC complementary segments on the two single-stranded DNA targets. The primer that binds to the forward strand is referred to as the \emph{forward primer} while the one that binds to the reverse strand is referred to as \emph{reverse primer}.
		\item Extension: The sample is heated to $72^{\degree}$C to enable the taq polymerase to extend the primers to form two complete copies of the original double-stranded DNA molecule.  
	\end{enumerate}
	Under ideal conditions, at the end of the Extension step of a cycle, the amount of target DNA doubles. This setting is illustrated in Figure~\ref{fig:pcr1}. However, due to several factors~\cite{booth2010efficiency} including the efficiency of denaturation, primer annealing affinity, polymerase binding strength, and others, the DNA content may not double during each cycle. 
For example, denaturation requires heating the sample to a higher temperature which by itself may cause oxidative and other damages to the DNA being amplified. The efficiency of denaturation is measured in terms of the concentration of viable single-stranded DNA present after heating.
	
	During the primer annealing stage, single-stranded DNA strings previously denatured can anneal back, therefore prohibiting access to the primer segments. The primer annealing efficiency is captured by the proportion of single-stranded DNA with bound primers. 
	
	When the polymerase binds to the DNA-primer complex it forms a potentially unstable tertiary complex in which the polymerase can disassociate in a stochastic manner. The polymerase binding efficiency is captured by the fraction of tertiary products in the assay. The tertiary complexes formed during the early stage of a cycle are more likely to result in complete double-stranded DNA compared to those formed in a later stage of the cycle, due to cycle timing issues. This effect is captured through what is known as the extension efficiency of PCR. 
	
	These effects jointly contribute towards the reduction of the average efficiency of DNA amplification, which goes down from the expected doubling factor to some value $<2$, usually written as $(1+ \eta)$, where $\eta$ is referred to as the cycle efficiency. The doubling of the target material at every cycle corresponds to $\eta =1$. At the end of $i$ cycles, a sample with concentration $x$ DNA strands is amplified to a sample with concentration $x(1+\eta)^i$. More precisely, the cycle efficiency depends on the cycle number. Consequently, a more accurate amplification model should use the factor $\eta_j$ for cycle $j$, so that the amplified concentration after the $i^{th}$ cycle reads as $x\cdot  \Pi_{j=1}^{i} (1+\eta_j)$. 
	It is also known that $\eta_j$ decreases with $j$, which may be attributed to the fact that the primers used for amplification are more and more integrated into the DNA products and that the efficiency of the polymerase decreases. At the same time, for a small number of cycles (usually $i \leq 10$) the DNA products are hard to detect. As a result, it is a common practice to run $30-40$ cycles of PCR, depending on the expected original concentration of the double-stranded DNA to be amplified. 
	
	Note that the polymerase can also be active at temperatures below $72^\degree C$, thereby initiating the extension process. However, the polymerase is nonspecific at lower temperatures and leads to amplification of nonspecific DNA strands. The high concentration of the stronger and more stable GC bonds in the DNA strands hinders effective denaturation at $96^\degree C$. Regions with high GC bond concentration also form secondary products that prevent primer bonding~\cite{differences}. These phenomena all jointly contribute to ``noise'' in the amplification PCR process which is not associated with the cycle efficiency. Additional sources of noise such as CCD thermal noise and shot noise can lead to a further decrease in the reliability of data points at low signal levels~\cite{platts2008real}.  
	
	Also, primers may fail to attach to the DNA if the corresponding DNA primer regions contain mutations (indels or point mutations). Since the error is caused by the actual DNA sample strand, and not the PCR process, this phenomenon should not be considered as part of the PCR noise model. The results of a simulation that studies the effect of mutations along the primer region on PCR amplification are described in Section~\ref{modeling}, using a collection of real genomic datasets retrieved from the GISAID database~\cite{gisaid}.
	
\subsection{Quantitative (real-time) RT-PCR}\label{sec:qrtpcr}
	
	{Quantitative Real Time PCR} (qPCR) is a technique used for precise analysis of viral and bacterial samples. As implied by its name, qPCR allows for the amplification process to be monitored in real-time. This is achieved by introducing fluorescent labels into the DNA products and recording the change in fluorescence with an increasing number of cycles (which also allows for estimating the number of cycles needed to detect an appropriate product). The result of a qPCR experiment is usually given in terms of an \emph{amplification curve} (an example of such a graph is shown in Figure~\ref{fig:3}, where real measurements are approximated by piecewise polynomial fragments of degree $\leq 10$). The amplification curve plots the normalized (relative) fluorescence $\Delta R_n$ against the cycle number. The fluorescence increases with the increase in the target genetic material with every cycle until the fluorescence saturates. The cycle number for which the fluorescence crosses the detection threshold (which can be defined in several ways) is referred to as the \emph{cycle threshold}, and denoted by $C_t$. Note that $C_t$ is inversely proportional to the concentration of the target material in the sample: A low $C_t$ value indicates a higher concentration of the sample we wish to detect, while a high $C_t$ value indicates a low concentration of the same or spurious amplification results. The slopes of the curves most often show very small variations with the concentration of the subject but may potentially be used as further indicators of the sample load.    
	
\begin{figure}[h]
		\centering
		\includegraphics[width=11.5cm]{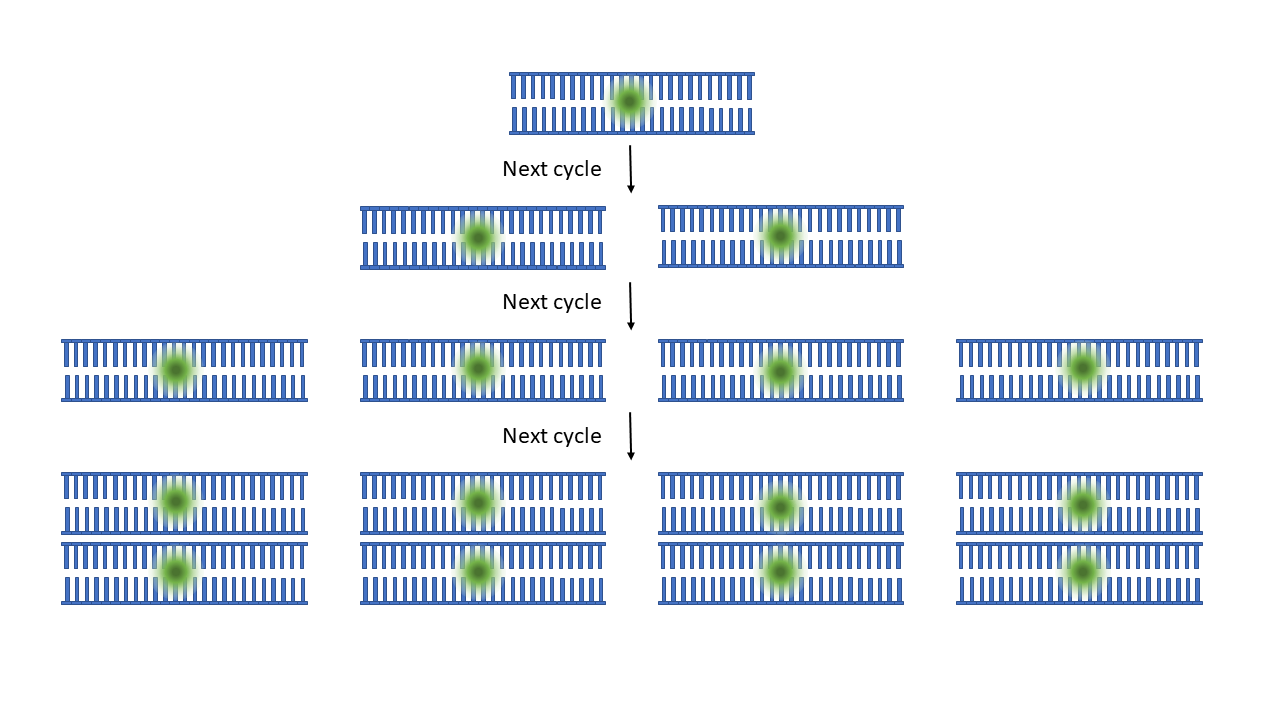}
		\caption{Dye-based qPCR: The dye attaches to the double-stranded DNA formed at the end of the extension stage and fluoresces. Thus, the fluorescence measured increases with the number of cycles.}  \label{fig:4}
	\end{figure}

\begin{figure}[h]
		\centering
		\includegraphics[width=12.5cm]{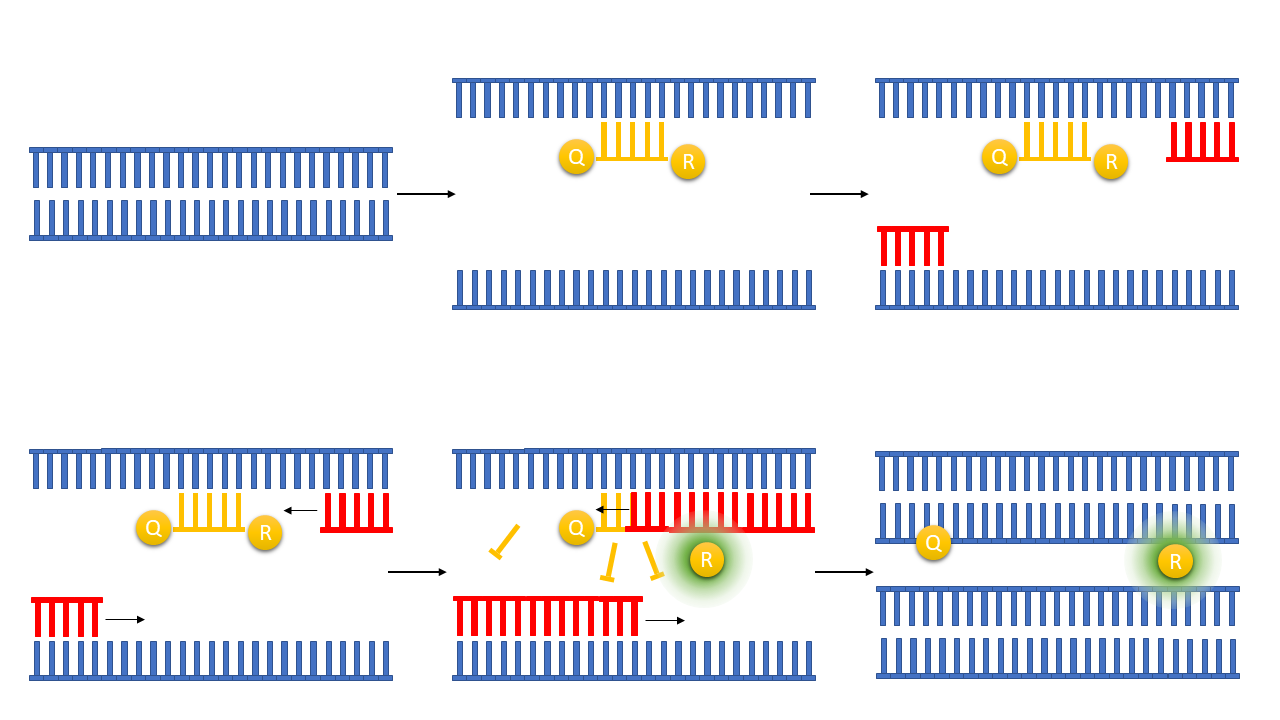}
		\caption{Probe-based qPCR: When DNA is denatured, a primer specific to the target DNA is attached to a single strand. The primers are then attached and extended by the polymerase. During the extension, the probes are cleaved and the reporter dye is no longer in the proximity of the quencher molecule, which enables it to fluoresce.} \label{fig:5}
	\end{figure}

\begin{figure}[h]
		\centering
		\vspace{-0.1in}
		\includegraphics[width=13cm]{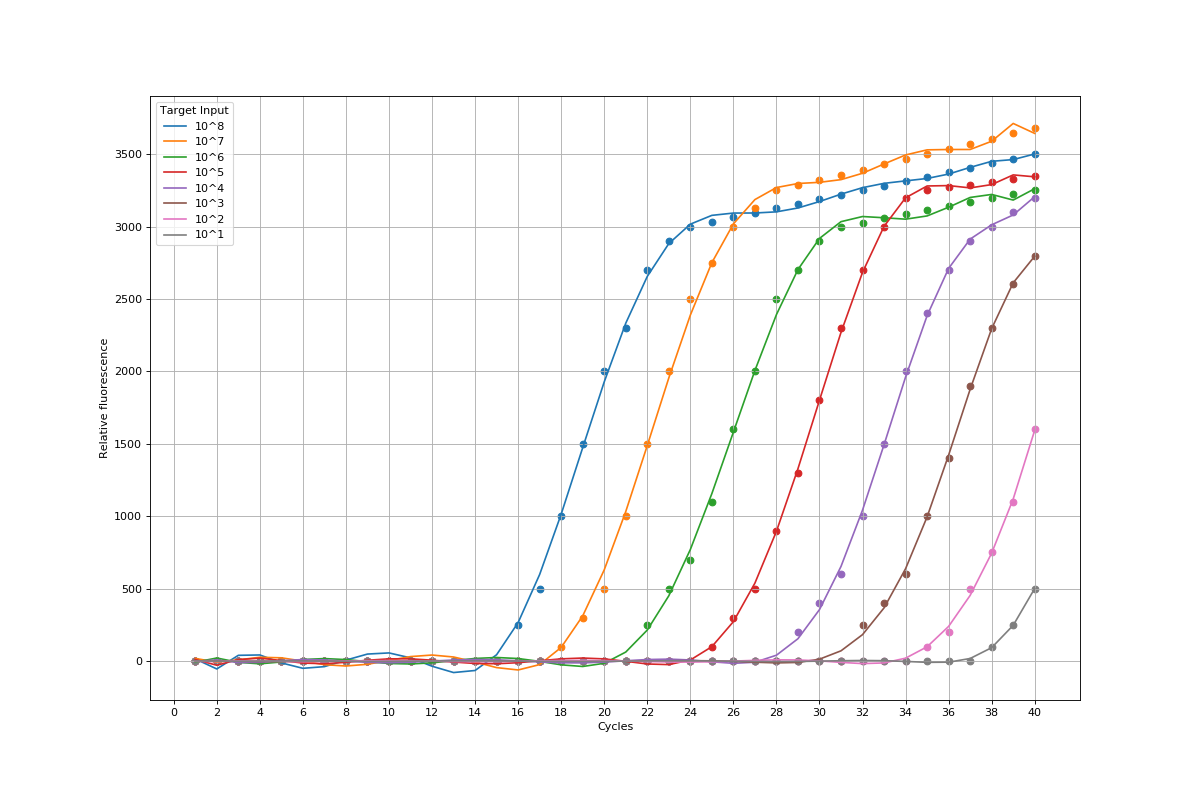}
		\caption{A typical amplification graph, plotting the relative fluorescence versus the number of PCR cycles for various input concentrations of the DNA sample. The dots represent actual fluorescent levels, while the curves represent a degree-$10$ polynomial approximation of the measurements. Since the solid curves are approximations, the fluorescence level for a small number of cycles can be negative, which is clearly not physically possible. Simple, yet less precise piecewise linear and quadratic curves will be described when discussing error models for real-time PCR. Also, note that the fluorescence saturates after roughly $35-40$ cycles which shows that models that use the final cycle fluorescence cannot distinguish viral loads. Another observation is that due to the stochastic nature or RT-PCR it usually takes around $5-10$ cycles to obtain visible fluorescence, independent of the viral load. Both of these features demonstrate the highly nonlinear relationship between the viral load and the fluorescence.} \label{fig:3}
	\end{figure}

Real-time or qPCR is usually performed using one of the following two approaches: 
	\begin{itemize}
		\item \textit{Dye-based qPCR.} The dye-based method uses dyes that \emph{only} fluoresce when bound to double-stranded DNA. Thus, at the end of each extension stage, the fluorescence increases (see Figure~\ref{fig:4}). The chemistry of the dyes used helps in distinguishing desired and undesired products. However, the dye-based method is often nonspecific, thereby inaccurately quantifying genetic material that is not of interest. As a result, this approach requires highly selective primers and other additional controls to provide accurate amplification curve results. 
		\item \textit{Probe-based qPCR.} In this technique, primers specific to the target DNA include two molecules, a fluorescent reporter dye and a quencher on its two ends. When the quencher is in close proximity to the fluorescent dye, the former molecule inhibits (quenches) the fluorescence of the latter. This is usually the case when the primer is not bound to the target (see Figure~\ref{fig:5}). However, when the primer is hybridized to the target and the polymerase extends the primer segment, the quencher and reporter separate out and the dye is cleaved and displaced. In its free form, it fluoresces which leads to detectable signals.    
	\end{itemize}
	
\begin{figure}[h]
		\centering
		\includegraphics[width=13cm]{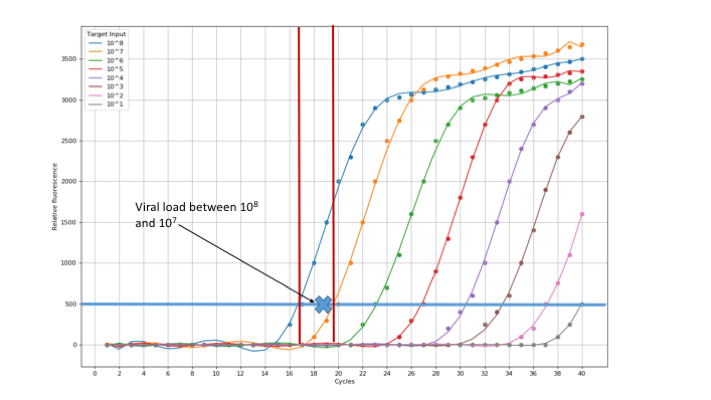}
		\caption{Amplification curves and quantization regions for the $C_t$ values. Given a number of amplification curves used for calibration in a specific lab, the quantization regions in this example are chosen based on the intersection of the fluorescence detection level $500$ and the calibration amplification curves. A $C_t$ value for a particular experiment is placed in the quantization region bounded by the two ``closest'' amplification curves used for calibration and their underlying $C_t$ values, or into the corresponding quantization bin. In this particular example, except for the quantization regions corresponding to the early and late cycles, the quantization regions are of nearly-uniform length. Note that the larger the $C_t$ value, the lower is the viral load. Also, if one were to only use the fluorescence levels observed at the final RT-PCR cycle (i.e., cycle number $\sim$40), the results would be noninformative with respect to the viral load as a strong saturation effect comes into existence.} \label{fig:quantex}
	\end{figure}  
		
\subsection{Amplification curves and the viral load}

\begin{figure}[h]
		\centering
		\includegraphics[width=12.5cm]{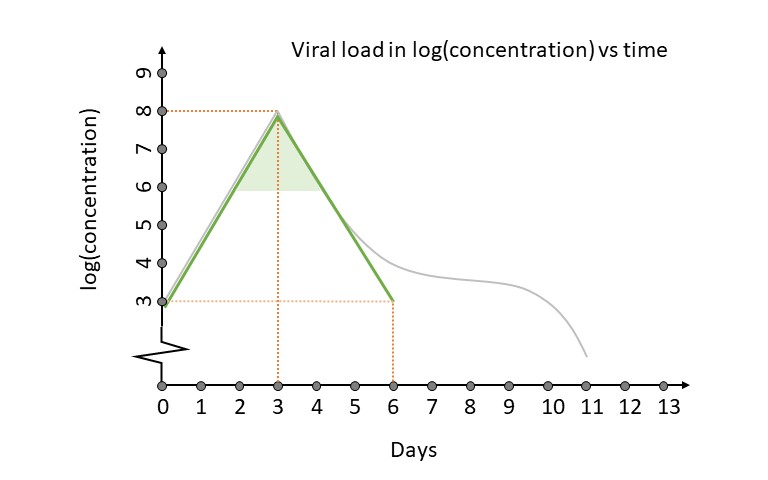}
		\caption{A typical viral load dynamics in an infected individual versus the time since infection. The viral load sharply spikes within the first three days of infection and then more gradually decreases. The nonlinear part of the viral load curve can be approximated by a linear component symmetric with respect to the linear component. This linear approximation will be used to determine the probability of heavy hitters, i.e., individuals who have an absolute viral load above $10^6$.} \label{fig:vload}
	\end{figure}

From Figures 7 and~\ref{fig:quantex}, and as already discussed in the previous section, it is clear that one can \emph{estimate upper/lower bounds} on the viral load of an individual by observing the $C_t$ value and the slope and saturation point of the amplification curve. It is important to point out that the viral load of individuals may vary up to five orders of magnitude, as shown in the recent study~\cite{goyal2020potency}. Viral loads in infected individuals tend to follow a ``typical'' inverted-V dynamics shown in Figure~\ref{fig:vload}. There, it can be seen that an individual tested $3-5$ days after the infection may have a viral load that is large enough to mask any other individual tested by the same test under the GT framework. This is a sensitive issue for SQGT schemes as the $C_t$ curves may have multiple interpretations: As an example, the same $C_t$ value may correspond to $10-100$ individuals tested $5-6$ days after infection or one individual tested $3$ days after infection. There are multiple possible ways to mitigate this problem: First, given that high viral loads very often positively correlate with observable disease symptoms~\cite{liu2020viral},\footnote{According to this study, among the set of infected patients, those who exhibited ``severe" symptoms had \textit{significantly lower} $C_t= C_{t}(\text{sample}) - C_t(\text{reference})$ values than those who exhibited ``mild'' symptoms.} asking individuals about symptoms before scheduling the tests (as is, for example, done at UIUC~\cite{uiuctesting}) allows one to determine if the individual should be group-tested or not. Another approach is to perform adaptive testing where samples with large viral loads are subjected to additional screenings, as is done in one of our proposed methods. Specialized testing strategies for pooled measurements with high viral loads can also be devised using heavy-hitter detection methods~\cite{indyk2013sketching}. 

As an abstraction, and only for our worst-case analysis we assume that each individual is represented by a viral load equal to the expected value over the tested population.
In this case, the test outcome can be translated into an interval in which the number of infected individuals lies. Hence, the assumption in this case is that one can convert $C_t$ values into a rough estimate of the number of infected people in the test. For probabilistic testing, we do not have to rely on such assumptions as the testing scheme itself can be easily adapted to handle heavy hitters. 

\section{Basics of GT} \label{bg:gt}
	
	In what follows, we provide concise overviews of all relevant GT schemes used or proposed for potential use for Covid-19 testing: (1) Classical nonadaptive and adaptive GT; (2) Nonadaptive SQGT; (3) Threshold GT; (4) Compressive sensing (CS); (5) Graph-Constrained GT. For all these methods, we describe their potential advantages and drawbacks and then proceed to introduce a new method, which we refer to as adaptive SQGT. Adaptive SQGT with a ``curve fitting''-based noise model appears to provide the theoretical state of the art GT results for qPCR test models and is the focus of our subsequent discussions.
	
	\subsection{Nonadaptive and adaptive GT}
	
	The assumptions are as follows: In a group of $n$ individuals, there are $d$ infected people. When a subset of people are tested, the result is positive (e.g., equal to $1$) if at least one person in the tested group is infected, else the test result is negative (e.g., equal to $0$). Such a testing scheme is referred to as \emph{binary}, as the outcomes take one of two values (see Figure~\ref{fig:comparison} (a)). GT aims to find the set of all infected people with the fewest number of binary tests possible and may use nonadaptive and adaptive tests. In the former case, all tests are performed simultaneously and the outcome of one test cannot be used to inform the selection of the individuals for other tests. In the adaptive setting, one can use multiple stages of testing and different combinations of individuals to best inform the sequentially made test choices. 
	
	When $d \ll n$, it is well-known that $\Omega (d \cdot \log (n/d))$ number of tests are required to find all infected individuals. Furthermore, it was shown in~\cite{d1982bounds} that for NAGT, at least $\Omega (d^2 \cdot \log (n)/ \log (d))$ tests are required. For the same parameter regime, there exist explicit nonadaptive schemes that require $\cO (d^2 \cdot \log (n/d) )$ tests to find the infected group~\cite{porat2008explicit}. A four-stage adaptive scheme that uses an optimal number of tests that meets the lower bound was recently described in~\cite{scarlett2019efficient}. Of special interest is the classical binary search result of~\cite{H72} which established an elegant adaptive scheme that differs from the information-theoretic limit only by an additive $\cO(d)$ term. 
	
	Despite the many proposed applications of this model to Covid-19 testing, it is obvious from the previous discussion that the GT measurement outcomes do not fully use the actually available qPCR results. One could argue that the fluorescence exceeding the detection threshold may correspond to the test outcome $1$, but clearly, significantly more information is available as the detection threshold depends on the concentration of the viral cDNA and hence the number of infected individuals. This motivates using a more quantitative GT approach, already introduced under the name of SQGT.
	
	\subsection{Nonadaptive SQGT}
	
	In SQGT, one is given a collection of thresholds $0=\tau_1<\tau_2<\cdots<\tau_r$, and the outcome of each test is an interval $(\tau_i,\tau_{i+1}],$ where $0 \leq i \leq r-1$. The outcome of an experiment cannot specify the actual number of infected individuals but rather provides a lower and upper bound on that number, $\tau_{i-1}$ and $\tau_{i}$, respectively. If $\tau_i=\tau_{i-1}+1$ for all values of $i$ and $r=d$, the scheme is referred to as additive GT, or the adder model~\cite{lindstrom1975determining,wolf1985born}. 
The two models are depicted in Figure~\ref{fig:comparison} (b) and (c). The additive test model described in~\cite{lindstrom1975determining} requires $2 \cdot (n/ \log \, n) $ tests to determine all possible infected individuals, for $0 \leq d \leq n$.

Another special SGT case of interest assumes that the test results are additive up to some threshold $\tau$ and after that, they saturate~\cite{d1984generalized} (see Figure~\ref{fig:comparison}). This model is of special interest for Covid-19 testing as it takes the warm-up/saturation information into account and, in addition, under a proper noise model, captures the fact that amplification graphs have different $C_t$ values determined by the concentration of the viral load (an hence the approximate number of infected individuals). Furthermore, one can argue that the RT-PCR fluorescence intensity information is inherently semiquantitative~\cite{EM14} as the fluorescence levels and $C_t$ values can be placed into bounded bins determined by the number of cycles. This observation is explained in more detail in the next section, along with new theoretical results pertaining to adaptive SQGT schemes with appropriate noise models.
\begin{figure}[h]
		\centering
		\includegraphics[width=11.5cm]{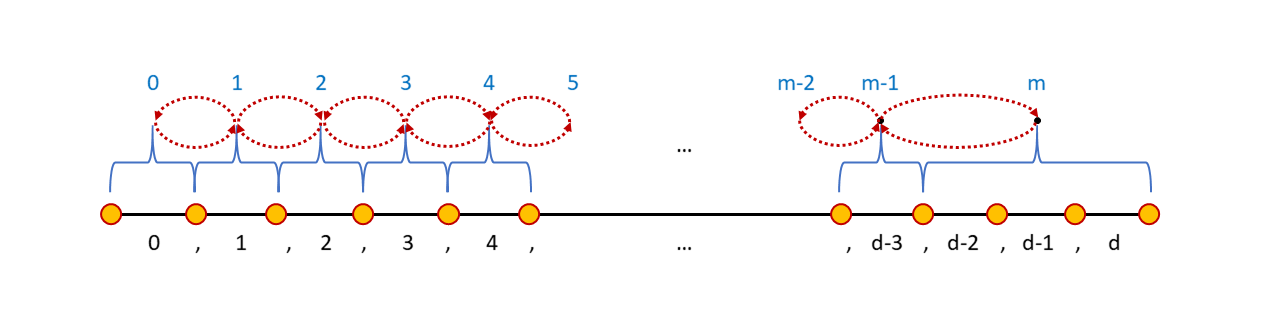}
		\caption{The birth-death noise model. Here, the assumption is that the $C_t$ value can be corrupted by noise only in so far that they can be mislabeled as belonging to intervals adjacent to the correct interval (except for the values falling into the first and last quantization region or bin).} \label{fig:nmodel}
	\end{figure}

	\subsection{Threshold GT}
	An extension of the GT problem was introduced by Damaschke in~\cite{damaschke2006threshold}: In this setting, if the number of defectives in any pool is at most the lower threshold $\ell>0$, then the test outcome is negative. If the number of defectives in the pool is at least the upper threshold $\emph{u}>\ell$, then the test outcome is positive. However, if the number of defectives in the pool is between $\emph{u}$ and $\ell$, the test outcome is arbitrary (i.e., either $0$ or $1$). Thus, the algorithms for Threshold GT are designed to handle worst-case adversarial model errors. Note that when $\ell=0$, and $\emph{u}=1$, Threshold GT reduces to GT. It is known that for nonadaptive threshold GT, $\cO (d\cdot g+2 \cdot (\log d) \log(n/d))$ tests (where $g =\emph{u}-\ell-1$) suffice to identify the $d$ infected individuals~\cite{cheraghchi2013improved}. 

The Threshold GT model is partly suitable for modeling the qPCR process, as the lower threshold can obviously assume the role of the fluorescence-based detection threshold, $\ell=C_t$; unfortunately, due to the saturation phenomena, a specialized choice for the upper threshold $\emph{u}$ does not allow one to accurately assess the number of infected individuals in the pool. The ``in-between'' threshold results also make the simplistic assumption that despite the observed fluorescence value being closer to the upper threshold, one can still call the outcome negative (and similarly for the small fluorescence levels and the lower threshold). 

\subsection{Compressive sensing}

In compressive sensing, the defectives are represented by nonnegative real-valued entries. Thus, quantitative GT represents a special instance of compressive sensing. Compressive sensing assumes that one is given an unknown vector $\textbf{x} \in \mathbb{R}^n$, in which only $d \ll n$ entries are nonzero. The vector $\textbf{x}$ is observed through linear measurements formed using a measurement matrix $M \in \mathbb{R}^{m \times n}$, leading to an observed vector $\textbf{y} = M\textbf{x}+\textbf{n}$, where $\textbf{n}$ is the measurement noise (usually taken to be Gaussian $\mathcal{N}(0, \sigma^2)$). For noiseless support recovery, $m=\cO(d \cdot \log\,\frac{n}{d})$ measurements are sufficient. For exact support recovery in the noisy setting, a signal-to-noise-ratio $S$ of $\Omega(\log \, n)$ is required for the same number of measurements as needed in the noiseless setting~\cite{aeron2010information}. Compressive sensing reconstruction is possible through linear programming methods or low-complexity greedy approaches~\cite{baraniuk2007compressive,tropp2004greed,dai2009subspace}. 

The recently proposed Tapestry method~\cite{ghosh2020compressed} combines group testing with compressive sensing and uses combinatorial designs (i.e., Kirkman systems) to construct the measurement matrix. However, the approach does not account for several practical features inherent to quantitative PCR. 
Although Tapestry proposes a model that involves multiplicative noise and converts it into additive noise through the use of a logarithmic function, it is still \emph{inherently linear:} Tapestry is based on a CS framework, which is additive and applies \emph{to viral loads}. But as seen from the previous discussion, PCR measurements report intersections of fluorescence level curves and a given threshold, and these values are nontrivial nonlinear functions of the viral load. Additionally, although the compressive sensing measurements used in their the work are assumed to correspond to $C_t$ values, no thresholding is used to model the actual practical process of generating the same\footnote{For many related questions arising in the context of group testing microarrays and quantized compressive sensing, the interested reader is referred to~\cite{dai2008compressive,dai2009comparative,dai2011information}).}. Also, this and all other methods do not account for the stochasticity of the PCR measurements and the fact that different lab protocols may lead to different $C_t$ values when presented with the same sample mixture.  
The CS methods in~\cite{ghosh2020compressed} rely on Gaussian assumptions for the cycle inefficiency exponent and do not take into account that the efficiency decays with the number of cycles and with the number of potential mutations in the primer regions (see also our analysis in~Section~\ref{bg}). As many other quantitative methods, it also appears vulnerable to heavy hitters. 

CS-based and many other proposed Covid-19 testing methods also do not take into account the fact that the number of RT-PCR machines/staff members is limited in virtually all test settings\footnote{PCR tests are performed on samples typically organized within $96$ wells, each of which can be used for one (group) test.}. The unavailability of arbitrary number of PCR machines inherently suggests using adaptive testing strategies. Adaptive quantitative testing schemes for Covid-19 were reported in~\cite{heidarzadeh2020two}. There, the same problem setup as in~\cite{ghosh2020compressed} is used to postulate an additive viral load model in the absence of noise. The new contribution of the work is a proposal for a two-stage testing scheme that bears a small resemblance to our methods in so far that we also propose two-stage adaptive pooling schemes. However, these techniques and the model used are different from ours since~\cite{heidarzadeh2020two} employs a combination of maximum likelihood and maximum-\textit{a-posteriori} estimators to determine the infected individuals in the second stage, while we employ zero-error GT and SQGT techniques to find \textit{all} infected individuals. Additionally, while~\cite{heidarzadeh2020two} reports the number of tests and conditional false positive and conditional false negative rates for the simulation experiments, we supplement our new tailor-made modeling and testing schemes with an in-depth theoretical analysis and performance guarantees. 
	
Nevertheless, there seem to be multiple advantages of CS methods for Covid-19 testing: One should be able, in principle, to recover not only the infected individuals but their viral loads as well (it still remains to be seen as such approaches are feasible as reported experiments use controlled concentrations of viral loads~\cite{ghosh2020compressed}). In particular, integer and nonnegative CS testing, along with quantized CS approaches can impose model restrictions on such testing schemes~\cite{dai2009comparative,DM09IntegerCS} to render them more suitable for the problem at hand. 
	
	\subsection{Graph-constrained GT}
	
	Let $\mathcal{G} = \{ V,E \}$ be a graph with vertex set $V$, $|V| =n$, and edge set $E$, representing a connected network of $n$ people out of whom $d$ are infected. In graph-constrained GT, vertices participating in the same test are restricted to form a path in the graph~\cite{cheraghchi2012graph}. This model is relevant as it can be adapted to require that only individuals that did not have contacts with each other are tested together (one only has to apply the problem to the dual of the contact graph used in Covid-19 testing). This allows us to identify individuals that fell ill in an ``independent'' fashion rather than through contact with each other. If $T(n)$ denotes the mixing time of the random walk on the graph, and $c=\frac{\Delta_{max}}{\Delta_{min}}$ is used to denote the ratio between the maximum degree and the minimum degree of the graph, then no more than $\cO( c^4 \cdot d^2 \cdot T^2(n) \log (n/d) )$ tests are required to find the set of infected vertices. For example, a complete graph ($T(n) =1, c=1$) requires no more than $\cO (d^2 \cdot \log(n/d))$ tests since it corresponds to the classical GT regime. Unfortunately, graph-constrained GT requires a significantly higher number of tests than classical GT methods as the tests are inherently restricted. As a result, despite the fact that this scheme is a natural choice for problems such as network tomography where these constraints need to be satisfied, it is not a proper choice for Covid-19 testing. Another ``community-constrained'' (although without an underlying interaction graph) was recently proposed in~\cite{nikolopoulos2020community} and is discussed in the next subsection. 	
	
\subsection{Community-aware GT}

Several lines of work have focused on what is now known under the name of \emph{community-aware GT}. In~\cite{nikolopoulos2020community, ahn2021adaptive}, the authors leverage correlations arising due to the presence of community structures to reduce the number of tests and increase the reliability of testing. More precisely, they assume that a community of $n$ members has $d \ll n $ infected individuals and that the population is partitioned into $F$ families. In the combinatorial infection model, it is assumed that $d_f$ families have at least one infected individual and that all the members of the remaining families are infection-free. An infected family indexed by $j$ is assumed to have $d^{(j)}$ infected members so that $d=\sum_{i=1}^{F} \, d^{(j)}$. The testing scheme can be succinctly described as follows: A representative individual from each family is selected uniformly at random. The representative community members are tested using either an adaptive or a nonadaptive GT algorithm. Family members whose representatives tested positive are tested individually. Members from the remaining families are tested together using either an adaptive or a nonadaptive GT scheme. The first approach proposed in~\cite{nikolopoulos2020community} did not account for inter-community interactions, but that issue was subsequently addressed in~\cite{nikolopoulos2020group}.

In a related line of work, the authors of~\cite{lin2020positively} establish how correlations in samples that arise due to the ordering of the tested individuals in a queue save in terms of pooled testing costs. In particular, the authors assume that in the first stage of Dorfman's testing scheme, the samples that are pooled and tested together are correlated. They model the correlation through the use of a random arrival process~\cite{lin2020positively}. The expected number of tests required to identify all infected individuals for their modified Dorfman scheme is compared against the expected number of tests required by the classical Dorfman testing scheme in which samples that are tested together are picked at random. The authors show that under certain conditions, the expected number of tests required by the modified Dorfman testing scheme does not exceed the expected number of tests required by the original scheme. Under additional conditions, the authors derive a closed form expression that captures the savings available for correlated samples. Furthermore,~\cite{lin2020positively} considers an underlying social contact graph, and  proposes an hierarchical agglomerative algorithm to identify individuals to be pooled together in the first stage of the modified Dorfman testing scheme. This line of work is closely related to the problem of identifying bursts of defectives, first introduced in~\cite{Colbourn1999Group} and analyzed for the case of a single burst.

We take a different approach to using community-structures for GT in so far that we suggest to \emph{quickly identify heavily infected (heavy hitter) families and then quarantine members of such communities.} To the best of our knowledge, this is the first GT problem formulation that also takes into account strategies for mitigating the spread of a disease, such as self-isolation. We address this question in the context of classical GT in Section~\ref{ongoing}.

Before proceeding with the original contributions, we remark that all the above GT techniques and scheduling models have \textit{probabilistic counterparts} in which each individual is assumed to be infected with the same probability $p$~\cite{Dor43} or members of different communities are infected with different probabilities, $p_i$, $i=1,\ldots,F$~\cite{hwang1975generalized}. The latter setting is especially important when prior information about the individuals is known (for example, their risk groups, potential symptoms etc). For an excellent in-depth review of these and some other GT schemes, the interested reader is referred to~\cite{AJS19}.

\section{AC-DC: new amplification curve based adaptive schemes - the probabilistic setting } \label{twostage}

Next, we introduce two adaptive SQGT schemes, one which is suitable for probabilistic testing and another one that is worst-case and nearly-optimal from the information-theoretic perspective. In the former case, considered in this section, a simple two-stage testing scheme is designed and analyzed with the goal of enabling practical implementations of adaptive SQGT. The results are described for two thresholds only, but a generalization is straightforward. This scheme also allows for incorporating heavy hitters into the testing scheme, which is of great practical relevance. In the worst-case, which is considered in the section to follow, the schemes extend the 
ideas behind Hwang's generalized splitting~\cite{H72} in two directions that lead to algorithms  using what we call \emph{parallel} and \emph{deep search}, respectively. In both settings, the outcomes of the first round of testing inform the choice of the composition of the test in the rounds to follow. The methods are collectively referred to as the AC-DC schemes, in reference to the use of the information provided by the amplification curve (AC) during the process of diagnostics of Covid-19 (DC). A relevant observation is that the worst-case adaptive schemes allow for using \emph{nonuniform amounts of genetic material from different individuals,} which may be interpreted as using \emph{nonbinary test matrices.}

\subsection{Practical adaptive AC-DC schemes} \label{sec:practical} We describe next a simple probabilistic two-stage AC-DC scheme that significantly improves upon the original single-pooling scheme of Dorfman and builds upon the SQGT framework. The underlying idea is to follow the same overall strategy as in the single-pooling scheme, but exploit the SQ information obtained in the first stage to perform better-informed testing in the second stage (i.e., dispense with individual testing of all individuals that feature in infected pools as part of the second stage).
	
	Consider a scenario where we have access to semiquantitative tests that return one of three values: If no individual featured in the test is positive, the test returns $0$. If between $1$ and $\tau$ individuals are positive, for some threshold $\tau \geq 1$, the test returns $1$. Finally, if more than $\tau$ individuals test positive, the test returns $2$. This scheme can be interpreted as follows: Suppose that $C_t$ is the observed cycle thresholds (defined in Section~\ref{sec:qrtpcr} for a particular test). If $C_t > c_1$ for some large threshold $c_1$, we say that the outcome is $0$ as the potential viral or viral-like contamination load is too small to claim the presence of an infected individual. If $c_2 \leq C_t \leq c_1$, we say that the output is $1$ and based on the average viral load, convert this into the maximum possible number of infected individuals $\tau$. If $C_t < c_2$, we say that the output is $2$ and that more than $\tau$ individuals in the pool are affected.
	
	For the new single-pooling AC-DC scheme, we assume that the population contains $n$ individuals, each of which is independently positive with some probability $p$ (Which can be easily determined based on regional infection rate reports: For example, at UIUC in September/October 2020~\cite{uiuctesting}, $p\simeq 0.05$), and proceed as follows:
	\begin{enumerate}
		\item \textbf{Stage 1:} Divide the $n$ individuals into $n/s$ disjoint pools $\cS_1,\dots,\cS_{n/s}$, each of size $s$;	
		\item \textbf{Stage 2:} \\
		\quad \quad  If a pool $\cS_i$ tests $0$, then immediately set the status of all individuals $\in\cS_i$ as ``negative''. \\	
		\quad \quad If a pool $\cS_i$ tests $1$, then apply a nearly-optimal zero-error nonadaptive group testing scheme to detect the $t$ infected individuals in $\cS_i$. (Such a testing scheme is simple to design: It suffices to sample a random binary matrix where all entries are i.i.d.\ according to some Bernoulli$(q)$ distribution, $0<q<1$. This is so since the resulting matrix will be a zero-error NAGT scheme with high probability provided the number of rows is large enough.) \\
		\quad \quad If a pool $\cS_i$ tests $2$, then test all individuals $\in\cS_i$ separately.
	\end{enumerate}
	Given the description above we can compute the expected number of tests per individual of the testing scheme, $T/n$, as a function of the probability of infection $p$, the first-stage pool size $s$, and the threshold $\tau$.
	Using the fact that the zero-error nonadaptive GT schemes we use in the second stage can be designed with $m(s,\tau)=c\cdot \tau^2\log(s/\tau)$ tests, we conclude that
	\begin{equation}\label{eq:expsqgt}
	\E[T/n]=\frac{1}{s}+p_1\cdot \frac{c\cdot \tau^2\log(s/\tau)}{s}+p_2,
	\end{equation}
where $p_1=\Pr[1\leq B(s,p)\leq \tau]$ and $p_2=\Pr[B(s,p)> \tau+1]$ denote the probability that a given pool tests $1$ and~$2$, respectively. Here, $B(s,p)$ stands for a binomial random variable with $s$ trials and success probability $p$.
	
\begin{figure}[h]
		\centering
		\includegraphics[width=0.9\textwidth]{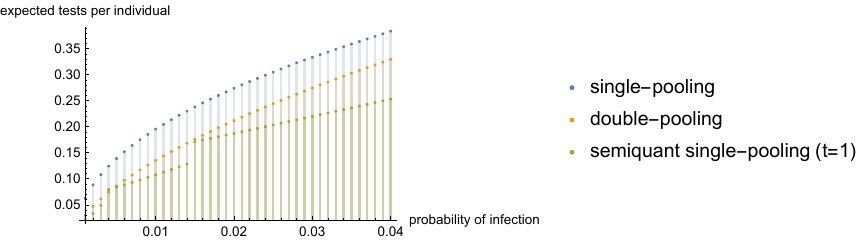}
		\caption{Comparison between the expected number of tests per individual required by Dorfman's single-pooling scheme~\cite{Dor43}, the Broder-Kumar double-pooling scheme~\cite{BK20}, and our semiquantitative single-pooling scheme.}
		\label{fig:comp-sp-dp-sqsp}
	\end{figure}

A particular case of interest pertains to setting $\tau=1$ in~\eqref{eq:expsqgt}. For a small probability of infection $p$, the optimal threshold $\tau$ is close to $1$ which justifies this choice. In this case, we have $p_1=s\cdot p(1-p)^{s-1}$ and $p_2=1-(1-p)^s-s\cdot p(1-p)^{s-1}$. Moreover, it is well-known that, for any $s$, there exists a simple (and optimal) zero-error nonadaptive scheme for finding $1$ defective among $s$ items using $m(s,1)=\lceil \log s\rceil$ tests (namely, set the $i$-th column of the test matrix to be the binary representation of $i$, i.e., use a Hamming code parity-check matrix for testing).
	Combining these observations together with~\eqref{eq:expsqgt}, we conclude that the expected number of tests per individual when $\tau=1$ equals
	\begin{equation}\label{eq:expsqgt1}
	\frac{1}{s}+p(1-p)^{s-1}\lceil \log s\rceil+1-(1-p)^s-s\cdot p(1-p)^{s-1}.
	\end{equation}	
On the other hand, the expected number of tests per individual for the basic single-pooling scheme~\cite{Dor43} is
	\begin{equation}\label{eq:expsp}
	\frac{1}{s}+1-(1-p)^s,
	\end{equation}
	and the expected number of tests per individual for the double-pooling scheme~\cite{BK20} is
	\begin{equation}\label{eq:expdp}
	\frac{2}{s}+p+(1-p)(1-(1-p)^{s-1})^2.
	\end{equation}	
	We compare the optimal expected number of tests per individual (as a function of $p$) achieved by our semiquantitative single-pooling scheme with $\tau=1$ (given in~\eqref{eq:expsqgt1}) and the single- and double-pooling schemes (given in~\eqref{eq:expsp} and~\eqref{eq:expdp}, respectively) in Figure~\ref{fig:comp-sp-dp-sqsp}. Semiquantitative single-pooling outperforms the other methods considered here as shown in the figure. This is in particular true for $p\leq 0.05,$ which we already pointed out corresponds to a practical parameter value. 
	
	There are two directions to further improve our scheme:
	\begin{itemize}
		\item We can easily extend the simple ideas presented above to obtain a semiquantitative version of double-pooling, and, more generally, multi-pooling schemes. The algorithm for this setting is summarized below:
	\begin{enumerate}
		\item \textbf{Stage 1:} Repeat Stage 1 of the previous semiquantitative scheme twice in parallel. We say an individual tests $(a,b)$ if its first pool tests $a$ and its second pool tests $b$.
		\item \textbf{Stage 2:} \\
		\quad \quad If an individual tests $(0,b)$ or $(a,0)$, immediately mark it as negative; \\
		\quad \quad If an individual tests $(1,1)$ or $(1,2)$, then apply a zero-error nonadaptive GT scheme for $\tau$ defectives to its first pool. \\
		\quad \quad If an individual tests $(2,1)$ or $(2,2)$, test them individually.
	\end{enumerate}
		\item We may also improve the performance of our semiquantitative scheme by introducing more (sufficiently small) thresholds $\tau_1<\tau_2<\cdots<\tau_\ell$ and extending the original idea in a natural way: If a pool has between $\tau_{i-1}$ and $\tau_i$ infected individuals, then apply a nearly-optimal zero-error NAGT scheme that detects $\tau_i$ infected to the pool in question.
	\end{itemize}

\subsection{Probabilistic SQGT with variable viral load} It is also simple to analyze how the SQGT scheme 
from the previous section performs when infected individuals may have either low or high viral loads, i.e., it is straightforward to account for heavy hitters. To this end, we consider a simplified model where each individual is independently infected and presents a low viral load at the time of testing with probability $p_{i1}$, or is infected and presents a high viral load at the time of testing with probability $p_{i2}$.
In particular, each individual is infected (regardless of her/his viral load) with total infection probability $p=p_{i1}+p_{i2}<1$.

As already explained, individuals with high viral load are problematic because, based on the SQ output of RT-PCR, pools featuring \emph{one} such individual may be mistaken for pools with \emph{several} infected individuals with low-to-average viral load.\footnote{This is not problematic for \emph{binary} group testing, where the test outcomes do not distinguish between one or several infected individuals in the pool.}
This phenomenon naturally leads us to consider the following modified version of the testing method studied in Section~\ref{sec:practical}: A test applied to a pool of individuals has outcome $0$ if there are no infected individuals in the pool, outcome $1$ if there exists \emph{exactly} one infected individual with \emph{low} viral load, and $2$ if either there exists more than one infected individual with low viral load, or at least one infected individual with \emph{high} viral load.
Therefore, as expected, individuals with high viral load obfuscate the test outcomes.

We consider now the SQGT scheme described in Section~\ref{sec:practical} with $\tau=1$ and under the heavy-hitter model. The probability that a pool of size $s$ contains exactly one infected individual with low viral load and zero individuals with high viral load (leading to test outcome $1$) is
\begin{equation*}
	s\cdot p_{i1} \cdot (1-p_{i1}-p_{i2})^{s-1}=s \cdot p_{i1} \cdot (1-p)^{s-1},
\end{equation*}
while the probability that the pool contains either more than one infected individual with low viral load or at least one individual with high viral load is
\begin{equation*}
	1-s \cdot p_{i1} \cdot (1-p_{i1}-p_{i2})^{s-1}-(1-p_{i1}-p_{i2})^s=1-s \cdot p_{i1} \cdot (1-p)^{s-1}-(1-p)^s.
\end{equation*}
Combining these observations with the reasoning from Section~\ref{sec:practical}, we conclude that the expected number of tests per individual as a function of $p_{i1}$ and $p_{i2}$ is given by
\begin{equation}\label{eq:expSQGTviral}
	\frac{1}{s}+s \cdot p_{i1} \cdot (1-p)^{s-1} \cdot \lceil\log s\rceil +1-s \cdot p_{i1} \cdot (1-p)^{s-1}-(1-p)^s,
\end{equation}
where $p=p_{i1}+p_{i2}$.

\begin{figure}[h]
	\centering
	\includegraphics[width=0.9\textwidth]{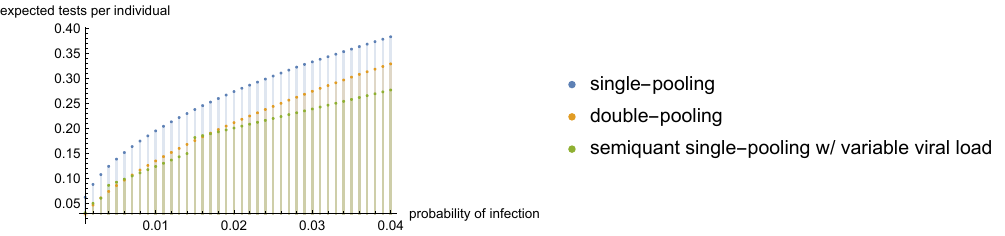}
	\caption{Comparison between the expected number of tests per individual required by Dorfman's single-pooling scheme~\cite{Dor43}, the Broder-Kumar double-pooling scheme~\cite{BK20}, and our semiquantitative semi-pooling scheme as a function of total infection probability $p$ with $p_{i1}=0.84p,\,p_{i2}=0.16p$.}
	\label{fig:comp-sp-dp-sqsp-viral16}
\end{figure}

\begin{figure}[h]
	\centering
	\includegraphics[width=0.9\textwidth]{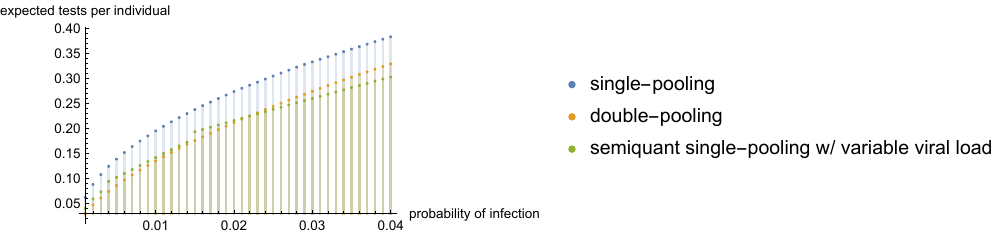}
	\caption{Comparison between the expected number of tests per individual required by Dorfman's single-pooling scheme~\cite{Dor43}, the Broder-Kumar double-pooling scheme~\cite{BK20}, and our semiquantitative semi-pooling scheme as a function of total infection probability $p$ with $p_{i1}=2p/3$ and $p_{i2}=p/3$.}
	\label{fig:comp-sp-dp-sqsp-viralthird}
\end{figure}

For fixed $p_{i1}$ and $p_{i2}$, it is easy to numerically minimize the expression above as a function of $s$ to find the optimal pool size for the scheme under consideration. Figures~\ref{fig:comp-sp-dp-sqsp-viral16} and~\ref{fig:comp-sp-dp-sqsp-viralthird} compare the expected number of tests per individual required by different schemes for different values of the total infection probability $p$ and the specific infection probabilities $p_{i1}$ and $p_{i2}$. The most practically relevant pair of parameters can be obtained from Figure~\ref{fig:comp-sp-dp-sqsp-viral16}, under the assumption that heavy hitters are individuals who have viral loads above $10^6$. Thus, by approximating the nonlinear portion of the viral load curve by a linear function, one can easily show that the probability that an infected individual is a heavy hitter is proportional to the area of the highlighted triangle, and approximately equal to $0.16$ (which is used in Figure~\ref{fig:comp-sp-dp-sqsp-viral16}). Note that the reduction in the number of tests increases with $p$, and for $p \sim 0.05$, which is a realistic infection rate, the savings compared to nonquantitative testing are larger than $5\%$.

Although we considered only an SQ single-pooling scheme in this section, these ideas can be easily extended to upgrade multi-pooling schemes with binary testing (such as the one from~\cite{BK20}) to exploit SQ test information under a variable viral load. This would allow one to further improve on the expected number of tests required by~\cite{BK20}.

\subsection{Adaptive SQGT with priors} The above described probabilistic setting can be generalized to account for different priors for different individuals by invoking the generalized binomial group testing scheme. Recall that Dorfman's scheme assumes that each individual has a probability $p$ of being infected, independent of everyone else. The set of individuals is partitioned into groups of size $s$, and each group is tested once. The group size $s$ is selected to minimize the number of expected tests. 

Hwang extended this setting in~\cite{hwang1975generalized} to account for the varying prior probabilities of infection. In this setting every individual $i \in \{ 1, 2, \dots, n\}$ is assumed to have probability $p_i$ of being infected, and thereby a probability of $q_i=1-p_i$ of not being infected. Clearly, in this generalized setting, a (possibly random) partitioning of the individuals into groups of equal sizes is no longer the optimal strategy to for the first round of testing.  

To find the optimal partition of the individuals in the generalized binomial setting, without loss of generality, one may assume that the individuals are reindexed so that $0 < q_1 \leq q_2 \leq \dots \leq q_n<1$. 

Given a subset of individuals to be tested, $\mathcal{G} \subseteq \{ 1, 2, \dots, n\}$, let $T(\mathcal{G})$ denote the expected number of tests required to find the set of infected individuals in $\mathcal{G}$ by first jointly testing all the individuals in the group $\mathcal{G}$ and then testing every member of $\mathcal{G}$ individually if the first test is positive. 
This number equals:
\begin{align}
    T(\mathcal{G}) = \begin{cases}
    1, &\text{ if } |\mathcal{G}| =1, \\
    1+ (1-\Pi_{j \in \mathcal{G}} q_j )|\mathcal{G}|, &\text{ otherwise}.
    \end{cases}
\end{align}
Now, let  $D(\mathcal{U}) =  \{ \mathcal{G}_1, \mathcal{G}_2, \dots, \mathcal{G}_k \}$ denote the optimal partition of the population $\mathcal{U}$ to be tested, where $|\mathcal{U}|=n$.
Let $C(\mathcal{U})$ denote the total expected number of tests required to run the two-stage testing procedure on the optimal partition. Clearly, $C(\mathcal{U}) = \sum_{i=1}^{k}  T(\mathcal{G}_i) $. Furthermore, let $\mathcal{U}_m$ denote the set of $m$ individuals with the highest probability of being infected, i.e., individuals indexed by $\{1,2, \dots , m \}$. The optimal partition and the corresponding expected number of tests can be found by solving the following optimization problems:
\begin{align}
    C(\mathcal{U}_m) = \min_{m-L-1 \leq i < m - \lceil (L +1)/2 \rceil} \{ T(\mathcal{U}_m - \mathcal{U}_i ) + C(\mathcal{U}_i) \}, 2 \leq m \leq n,
\end{align}
where $L$ denotes the size of the largest group/part in $D(\mathcal{U}_{m-1})$. At step $m$ of the optimization procedure, the probabilities $\{p_j \}_{j \leq m}$ and the previously computed $C(\mathcal{U}_j)$  are used to determine $D(\mathcal{U}_m)$ and the expected number of tests $C(\mathcal{U}_m)$ of the population $\mathcal{U}_m$.  As a consequence of the structure of the program, one has the following property for the optimal partition: If individuals $i$ and $j$, $j >i$, are in the same group, then all individuals $k$ such that $i<k<j$ are in the same group as well (a simple induction argument can be used to prove this claim).  

Next, assume that the population has only two types of individuals: $m$ individuals that have a high $p_1$ probability of infection, and the remaining $n-m$ individuals that have a low $p_2 \ll p_1$ probability of infection. Exploiting the structure of the optimization program and the fact that only two types of individuals are present in the test pool, the optimal number of tests needed to find the set of infected individuals using the two-stage procedure equals:
\begin{align*}
    \min_{s_1,\,s_2,\,x,\,y\, \in \mathbb{N}\cup \{ 0 \}} \left[ \frac{m-x}{s_1} \left( 1 + (1-q_1^{s_1}) s_1 \right) + \frac{n-m-y}{s_2} \left(1 + (1-q_2^{s_2}) s_2 \right) + \mathbf{1}_{xy>0} + (1-q_1^x q_2^y)(x+y) \right],
\end{align*}
where $x$ and $y$ represent the number of individuals that have high and low probabilities of infection, respectively, and are tested together, while $s_1, s_2$ are the sizes of the groups used for testing individuals with high and low probabilities of infection, respectively. The optimization allows for at most one heterogeneous group of size $x+y$. The average number of tests required to find all the infected individuals in this heterogeneous group is given by $\mathbf{1}_{xy>0} + (1-q_1^x q_2^y)(x+y)$. The remaining $m-x$ individuals who have $p_1$ probability of infection are divided equally into groups of size $ s_1$, where each group requires, on average, $ 1 + (1-q_1^{s_1}) s_1$ tests to determine the set of all infected individuals. Similarly, the $n-m-y$ remaining individuals who have $p_2$ probability of infection are divided equally into groups of size $ s_2$, where each group requires, on average, $ 1 + (1-q_2^{s_2}) s_2$ tests to determine the set of all infected individuals.

\subsection{Lower bounds for nonadaptive probabilistic SQGT} We conclude our exposition in this section by presenting a theoretical result that establishes lower bounds for nonadaptive probabilistic GT that may be used to assess the quality of our adaptive schemes. For this purpose, we adapt an argument by Aldridge~\cite{Ald18} for arbitrarily small error probability under a constant probability of infection.
	More precisely, we consider a setting where each test has $m+1$ outcomes for some $m \geq 1$: The outcome of a test is either $i$ if there are exactly $i$ infected individuals for $i<m$, and $\geq m$ otherwise. This corresponds to the setting introduced in~\cite{d1984generalized} which provides the most informative type of measurements one can expect from the SQGT framework using the amplification curve information. This model accounts for the saturation limit for each test, dictated by $m$, which is a phenomenon observable from the amplification curve. Moreover, as before we assume that each individual in the population of size $n$ is infected independently with some constant probability $p>0$.
	We show the following.
	\begin{thm}\label{thm:lb}
		For every $m$ and constant $p>0$ there exists a constant $\eps(m,p)>0$ such that, under the setting described above, nonadaptive testing requires at least $n/m$ tests to achieve error probability less than $\eps(m,p)$ in a population of size $n$.
	\end{thm}
	In contrast, for $m=2$, our two-stage scheme uses significantly fewer than $n/2$ tests provided $p$ is not very large.
	
	Proving Theorem~\ref{thm:lb} follows by a simple adaptation of an approach by Aldridge~\cite{Ald18}, who showed that individual testing is required in order to achieve arbitrarily small error in regular nonadaptive probabilistic group testing (which corresponds to $m=1$).
	First, given any nonadaptive testing scheme, we may without loss of generality remove all tests with $m$ or fewer elements, along with all individuals who participate in those tests. This does not affect the lower bound. Then, we show that there are no nonadaptive testing schemes with an arbitrarily small error where every test includes at least $m+1$ individuals.
	Combining these two observations immediately yields Theorem~\ref{thm:lb}.
	
	For an individual $i$, let $x_i$ denote its infection status.
	Call an individual $i$ (regardless of its infection status) \emph{disguised} if every test $t$ in which it participates contains at least $m$ other individuals which are infected.
	If $i$ is disguised, then changing $x_i$ from $0$ to $1$, or vice-versa, does not change the outcome of the testing scheme.
	As a result, we can do no better than guess $x_i$, and we will be wrong with probability at least $\min(p,1-p)$.
	To finalize the argument, it suffices to show there is a disguised individual with constant probability.
	
	Let $D_i$ denote the event that individual $i$ is disguised, and let $D_{t,i}$ denote the event that individual $i$ is disguised in test $t$.
	Since the $D_{t,i}$ are increasing events\footnote{If $D_{t,i}$ holds and the set of infected individuals is expanded, then $D_{t,i}$ continues to hold under this expanded set}, the Fortuin-Kasteleyn-Ginibre (FKG) inequality~\cite{fortuin1971correlation} implies that
	\begin{equation}
	\Pr[D_i]\geq \prod_{t:x_{t,i}=1}\Pr[D_{t,i}],
	\end{equation}
	where $x_{t,i}$ indicates whether individual $i$ participates in test $t$.
	Moreover, we have
	\begin{equation}
	\Pr[D_{t,i}]=\Pr[B(w_t-1,p)\geq r],
	\end{equation}
	where $w_t=\sum_{i=1}^n x_{t,i}$ is the weight of test $t$.
	
	Let
	\begin{equation*}
	L_i=\log\left(\prod_{t:x_{t,i}=1}\Pr[D_{t,i}]\right)=\sum_{t:x_{t,i}=1}\log \Pr[D_{t,i}]=\sum_{t=1}^T x_{t,i}\log \Pr[D_{t,i}],
	\end{equation*}
	where $T$ denotes the total number of tests, which we assume satisfies $T/n<1$.
	Then, it suffices to show that there exists some $i^\star$ with $L_{i^\star}>c$ for some constant $c$ independent of $n$.
	Let $I$ be uniformly distributed over $\{1,2,\dots,n\}$, and let $\overline{L}=\E[L_I]$.
	We have
	\begin{align*}
	\overline{L}&=\frac{1}{n}\sum_{i=1}^n \sum_{t=1}^T x_{t,i}\log \Pr[D_{t,i}]\\
	&=\frac{1}{n}\sum_{t=1}^T w_t\log \Pr[D_{t,i}]\\
	&\geq \min_{t=1,\dots,T} w_t\log \Pr[B(w_t-1,p)\geq m]\\
	&\geq\min_{w\geq r+1} w\log \Pr[B(w-1,p)\geq r]=:L^\star,
	\end{align*}
	where the second equality follows from the fact that $\Pr[D_{t,i}]$ is the same for every $i$ such that $x_{t,i}=1$, and in the first inequality we use the assumption that $T/n<1$.
	It is immediate that there exists some $i^\star$ with $L_{i^\star}\geq \overline{L}$, which implies that $\Pr[D_{i^\star}]\geq 2^{L^\star}$.
	Therefore, the error probability of the testing scheme is at least $\eps(m,p)=\min(p,1-p)\cdot 2^{L^\star}$.
	Noting that $L^\star$ does not depend on $n$ and is bounded from below for any $m$ and $p$ (since $\lim_{w\to\infty} w\log \Pr[B(w-1,p)\geq m]=0$) concludes the proof.	

\section{AC-DC Schemes: Worst-case model analysis } \label{sec:worst}
	
As before, we assume that we are given a set of $n$ samples with at most $d$ infected individuals. Our goal is to minimize the number of tests needed to identify \emph{all infected individuals} and we do not impose any restrictions on the ``simplicity'' of our scheme. As a result, we consider a generalization of the model described in the previous section which allows for more than three test outcomes. 

For simplicity, as well as for practical reasons\footnote{As we quantize the $C_t$ values or the phase transition thresholds according to equally spaced cycle numbers}, we focus on equidistant thresholds but allow for warm-up/saturation effects. We refer to this model as the saturation GT scheme.

Let $\tau, m \in \mathbb{Z}^{+}$ represent the distance between the thresholds and the number of thresholds, respectively. 

Denote the outcomes of the test by a nonnegative integer $t \leq m$. Then,
	\begin{align}\label{eq:tret}
	t=\begin{cases}
	0, &\text{ if every sample in the test is negative},\\
	1, &\text{ if the number of infected individuals is between $1$ and $\tau$}, \\
	2, &\text{ if the number of infected individuals is between $\tau+1$ and $2\tau$}, \\
	\vdots  & \vdots \\
	m-1, &\text{ if the number of infected individuals is between $(m-2) \tau +1$ and  $(m-1)\tau$, or} \\
	m, &\text{ if the number of infected individuals is at least $(m-1)\tau + 1$.}
	\end{cases}
	\end{align}
	We seek to identify $d$ infected individuals from a population of size $n$ given that each test returns a value in~(\ref{eq:tret}). We refer to this problem as the $(n,d)$ adaptive SQGT problem or the $(n,d)$-ASQGT problem for short. 
	
	Another way of looking at (\ref{eq:tret}) is that if the collection of samples tested contains $d'$ infected individuals, then the output of the test is $\lceil \frac{d'}{\tau} \rceil$ when $d' \leq m \tau$ and $m$ otherwise. Note that for every test there are $m+1$ possible outcomes and the output of a test tells us roughly (within at most $\tau$) how many total infected samples are part of the tested pool of samples. 
	
	\begin{remark} Note that this model differs from the model introduced in Section~\ref{twostage} since as $m$ increases the widths of our threshold remain the same whereas in Section~\ref{twostage} the widths changes as the number of thresholds increases. Despite this difference, and as will be discussed in Example~\ref{ex:widths}, the ideas discussed here are applicable to the case where the widths of the thresholds are nonuniform.
	\end{remark}  
	
	Let $2^{\beta} = m+1$. Motivated by practical applications, we will be interested in the case where $\beta = \cO(1)$.
	Our main results are two algorithms, which we refer to as \textit{\textbf{parallel search}} and \textit{\textbf{deep search}}. Parallel search is applicable for the setting $d > \beta$. In Lemmas~\ref{lem:ps1} and \ref{lem:ps2}, we show that using parallel search it is possible to efficiently identify from a set of pools (each of size $s=2^{\alpha}$ and large enough to contain at least $\beta$ infected individuals) a set of $\beta$ defectives using at most $\alpha$ tests. Note that as a first-step simplification, one may think of $n$ being approximately equal to $d\cdot 2^{\alpha}$; the notation involving $\alpha$ is chosen to enable a comparison between our SQGT search scheme and the well-known splitting approach by Hwang~\cite{du2000combinatorial}. Deep search, discussed in Lemma~\ref{lem:ds} and applicable for the setting $d < \beta$, shows that it is possible to identify all $d$ infected individuals using roughly $\frac{d \cdot \alpha}{\beta - \log(\beta)}$ tests. Our main result is Algorithm~1, which for $d = \Omega(n)$ shows that one can identify $d$ infected individuals using at most $\frac{d}{\beta} \cdot \left( \alpha + 3 + \log \beta\right)$ tests. These results show that adaptive SQGT roughly provides $\beta$-fold savings in the number of tests when compared to classical adaptive GT. Furthermore, they differ from the information-theoretic lower bound (as it applies to ASQGT) of Lemma~\ref{lem:lb} by $\cO(\frac{d}{\beta})$ tests. It remains an open problem to identify whether it is possible to solve the $(n,d)$-ASQGT problem using fewer tests.
	
	We start with the following obvious claim, which allows us to restrict our attention to the case where $\tau=1$ and simplifies the problem at hand.
	
	\begin{claim}\label{cl:gtw} Let $\cG$ be the set of test subjects and suppose that there are at most $d$ infected individuals within this group. Let $\cP^{(1)}$ be a pool formed by taking one sample from each individual in $\cG$ and let $\cP^{(w)}$ be a pool formed by taking $w$ samples from each individual in $\cG$. Let $t^{(1)}$ be the output of testing $\cP^{(1)}$ under the setup $(m, \tau) = (m, 1)$ and let $t^{(w)}$ be the output of testing $\cP^{(w)}$ under the setup $(m, \tau) = (m, w),$ according to (\ref{eq:tret}). Then, $t^{(1)} = t^{(w)}$.
	\end{claim}
	
	Next, we present a lower bound (i.e., information-theoretic or counting lower bound) on the number of tests necessary to solve the $(n,d)$-ASQGT problem. The result follows from a simple counting argument and is consistent with the result from Claim~\ref{cl:gtw}, as it does not depend on the width $\tau$ of the threshold. 
	
	\begin{lemma}\label{lem:lb} Let $n=(2^{\alpha}+1) \cdot d + 2^{\alpha} \cdot \delta + \Delta$, where $\alpha, \delta, \Delta$ are integers, $\delta < d$, and $\Delta < 2^{\alpha}$. Then, the number of tests $L(n,d,m)$ needed to identify the infected individuals is bounded as:
		$$L(n,d,m) \geq \frac{d}{\beta} \cdot (\alpha + 1).$$ 
	\end{lemma}
	\begin{IEEEproof} The number of ways to select at most $d$ infectives in a group of $n$ individuals is $\sum_{i=0}^d \binom {n}{i}$. Thus, we have
		\begin{align*}
		L(n,d,m) \geq \log_{m+1} \left(\sum_{i=0}^d \binom {n}{i} \right) \geq& \log_{m+1} \binom {n-d+d}{d} \\
		\geq& \log_{m+1} \left( \frac{n-d + d}{d} \right)^d \\
		\geq& d  \cdot \log_{m+1} \Big( 2^{\alpha} \big( 1 + \frac{\delta}{d} + \frac{\Delta}{2^\alpha d} + \frac{1}{2^{\alpha}} \big) \Big)\\
		\geq& \frac{d \cdot \alpha}{\beta} + \frac{d}{\beta} \cdot \log_{2} \Big( 1 + \frac{2^{\alpha} \cdot \delta + \Delta + d}{2^{\alpha} d} \Big) \\
		\geq& \frac{d}{\beta} \cdot \left( \alpha + 1\right).
		\end{align*}
	\end{IEEEproof}	
The next example illustrates a simple approach for addressing the ASQGT problem, and motivates the analysis that follows.
	
	From here on, we write $[[x]]=\{{0,1,\ldots,x-1\}}$ and $[x]=\{{1,\ldots,x\}}$.
	
	\begin{example}\label{ex:motivate} Suppose that we are given a collection of $n$ individuals with exactly $d$ infected subjects. 
	We start by randomly partitioning the set of $n$ individuals into $d$ groups each of size $s=\frac{n}{d} = 2^{\alpha},$ where we assumed for simplicity that $d | n$. The expected number of infected individuals in each group is $1$.
		
		Denote the $d$ groups or pools by $\cG_0, \cG_1, \ldots, \cG_{d-1}$; all groups have the same size and from this point on, for simplicity, assume that each group contains \emph{exactly one} infected subject. For $i \in [[d]]$ we proceed as follows. We partition $\cG_i$ into $2^{\beta}$ groups of equal size and denote the subgroups as $\cG_i^{(0)}, \cG_i^{(1)}, \ldots, \cG_i^{(2^{\beta}-1)}$. Under this setup, there exists exactly one index $j^{\star}$ such that the number of infected individuals in $\cG_i^{(j^{\star})}$ equals to one, and every other group $\cG_i^{(j)}$, $j \in [[2^{\beta}]] \setminus j^{\star}$ is free of infected individuals. 
		
		Next, we form a new set of pools, which we denote by $\cP_i$, $i \in [[d]],$ comprising $k$ replicas of the samples in $\cG_i^{(k)}$, for all $k=0,\ldots,2^{\beta}-1$. Let $t_i$ denote the output of the semi-quantitative test described in (\ref{eq:tret}) after the pool $\cP_i$ is tested. 
		Then, it is straightforward to observe that the outcome $t_i$ is $j^{\star},$ and hence we can identify the group which contains the one single infected individual using only one \emph{nonbinary outcome} test. We repeat this procedure for each group $\cG_i,$ $i \in [[d]]$, partitioned into subgroups.
		It can be hence seen that it is possible to identify $d$ infected individuals using only $d\frac{\alpha}{\beta}$ tests assuming each of the $d$ groups of size $2^{\alpha}$ each contain \emph{exactly} one infected subject. 
	\end{example}
	
	To make this argument rigorous, we need to account for the fact that not every group will have exactly one infected individual. In this case, upon creating the subpools we have to recursively test them until we identify a prescribed number of infected individuals. In fact, the approach from the previous example is a special case of what we refer to as \emph{deep search}, described in Lemma~\ref{lem:ds}. The resulting algorithm is summarized in Algorithm~1, and it requires roughly an additional factor of $\cO(\frac{d}{\beta})$ tests compared to the information-theoretic lower bound.
	
	\subsection{Parallel search}\label{subsec:parallel}

We start by introducing some useful notation. Suppose that $\cG'$ is a subgroup of individuals to be tested and that the outcome of a test governed by (\ref{eq:tret}) is $t$. In this case, we say that $\cG'$ is a $t$-infected group. When referring to an ordered collection of groups $(\cG_0, \cG_1, \ldots, \cG_{g-1})$, we say that the collection is a $(t_0, t_1, \ldots, t_{g-1})$-infected group if $t_0 \geq t_1 \geq \cdots \geq t_{g-1}$ and $\cG_i$ is a $t_i$-infected group, for $i \in [[g]]$. We also say that $(\cG_0, \ldots, \cG_{g-1})$ is a $\beta$-\textit{\textbf{minimal group}} if $\sum_{j=0}^{g-2} t_j < \beta$, but $\sum_{j=0}^{g-1} t_j \geq \beta$. 

The following lemma constitutes the key component of one of our approaches to solving the $(n,d)$-ASQGT problem. We refer to the procedure described in the proofs of the next two results as parallel search. 

In the first lemma below, we make the simplifying assumption that a group is $\beta$-minimal and $g = \beta$. Afterward, in Lemma~\ref{lem:ps2} we consider the case when $g < \beta$.

	\begin{lemma}\label{lem:ps1} Let $\alpha$ and $\beta$ be positive integers. Suppose that $(\cG_0, \cG_1, \ldots, \cG_{\beta-1})$ is a $\beta$-minimal group, where $g = \beta$, and that each group has size at most $2^{\alpha}$. Then, we can identify $\beta$ infected individuals in the group using at most $\alpha$ tests.  
	\end{lemma}
	\begin{IEEEproof} We prove the result by induction on $\alpha$, where $2^{\alpha}$ as before is the size of each subgroup. Recall that under this setup $t_0=t_1=\cdots=t_{g-1}=1$ and $g = \beta$. 
			
		First, consider the case $\alpha=1$, for which we have $\beta$ $1$-infected groups of individuals and each group has size $2$. For shorthand, denote the $\beta$ infected groups as $\cG_{0}, \cG_1, \ldots, \cG_{\beta-1}$. From these $\beta$ groups, we form a ``super-pool'' of samples which contains a total of $2^0 + 2^1 + 2^2 + \cdots + 2^{\beta-1} = 2^{\beta}-1$ samples. More precisely, for $i \in [[\beta]]$, the super-pool contains $2^{i}$ samples from one individual $\in \cG_i$. Since $t_0 = t_1 = \cdots = t_{\beta-1}=1$ and $\tau=1$, according to (\ref{eq:tret}) the output returned after testing this super-pool of samples is a number $t$ between $0$ and $2^{\beta}-1$. Let $b_0, b_1, \ldots, b_{\beta-1}$ be the binary representation of the number $t$. It is straightforward to verify that $b_i = 1$ then the individual selected from $\cG_i$ is infected. Otherwise, if $b_i = 0$, then the above described individual is not infected, which implies the other individual (the one not tested) in group $\cG_i$ is infected. Thus, we conclude the statement in the lemma holds when $\alpha = 1$.
		
		For the inductive step, assume that the statement holds when the group size is at most $2^{\alpha'}$ and consider the setup where the group size is $2^{\alpha} = 2^{\alpha'+1}$. We follow the same approach as described for $\alpha=1$ for creating super-pools. Under this setup, we have $\beta$ $1$-infected groups $\cG_{0}, \cG_1, \ldots, \cG_{\beta-1},$ each of size $2^{\alpha' + 1}$. For $i \in [[ \beta]]$, let $\cQ_i \subseteq \cG_i$ be a subset of $\cG_i$ of size $2^{\alpha'}$. Next we construct a super-pool that contains $2^{i}$ samples from each individual in $\cQ_i,$ $i \in [[\beta-1]]$. Let $t$ denote the output of testing this super-pool according to (\ref{eq:tret}), where $b_0, b_1, \ldots, b_{\beta-1}$ is the binary representation of $t$. As before, if $b_i = 1$, then $\cQ_i$ has a single infected individual. Otherwise, if $b_i=0$, then there is an infected individual in the set $\cG_i \setminus \cQ_i$ which also has size $2^{\alpha'}$. For $i \in [[\beta]]$, let $\cG_i' = \cQ_i$ if $b_i=1$ and otherwise, if $b_i=0$, set $\cG_i' = \cG_i \setminus \cQ_i$. Then, $(\cG_0', \cG_1', \ldots, \cG_{\beta-1}')$ is a $(1,1,\ldots,1)$-infected group and we can apply the inductive hypothesis to $(\cG_0', \cG_1', \ldots, \cG_{\beta-1}')$.
	\end{IEEEproof}
	
	For the case $g < \beta$, we use a similar partitioning idea to identify at most $\beta$ subgroups which satisfy the conditions in the lemma. The difference between the approaches is that for $g < \beta$ the number of samples added into the pool is dictated by a mixed-radix representation (in which the numerical base varies from position to position) rather than a binary representation. For simplicity, we assume from now on that $\beta$ is an even integer although the results hold for odd integers as well.
	
	\begin{lemma}\label{lem:ps2} Let $\alpha, \beta,g$ be positive integers such that $g < \beta$. Suppose that $(\cG_0, \cG_1, \ldots, \cG_{g-1})$ is a $\beta$-minimal group and that each group has size at most $2^{\alpha}$. Then, we can identify $\beta$ infected individuals using at most $\alpha$ tests.  
	\end{lemma}
	\begin{IEEEproof} We begin with the following claim which we find useful in our subsequent discussion.
		
		\begin{claim}\label{cl:maxrep} Suppose we are given a sequence $(t_0, \ldots, t_{g-1}) \in [[\beta+1]]^g,$ where $g < \beta$, and the values $t_0 \geq t_1 \geq \cdots \geq t_{g-1}$ are such that $\sum_{j=0}^{g-1} t_j \geq \beta$, but $\sum_{j=0}^{g-2} t_j < \beta$. Furthermore, let $(n_0, \ldots, n_{g-1}) \in [[t_0+1]] \times [[t_1+1]] \times \cdots \times [[t_{g-1}+1]]$. Then, the number of different choices for $(n_0, \ldots, n_{g-1})$ is at most $2^\beta$.
		\end{claim}
		\begin{IEEEproof}[Proof of Claim~\ref{cl:maxrep}] First, consider the case $g \leq \frac{\beta}{2}+1$. Since $\sum_{j=0}^{g-2} t_{j} < \beta$, it follows that $t_{g-2} \leq \frac{\beta}{g-1}$ and, from the assumptions stated in the claim, $t_{g-1} \leq t_{g-2} \leq \frac{\beta}{g-1}$. The total number of possibilities for $(n_0, \ldots, n_{g-1})$ restricted to the first $g-1$ components is maximized when $t_0 = t_1 = \cdots = t_{g-2}$, which implies that the total number of possible choices for the $i$-th component of the sequence when $i \in [[g-1]]$ equals $\frac{\beta}{g-1} + 1$. Therefore, the total number of possibilities for the constrained sequences is
			\begin{align*}
			\left( \frac{\beta}{g-1} + 1 \right)^{g} \leq 3^{\beta/2 + 1},
			\end{align*}
			which follows since $\left( \frac{\beta}{g-1} + 1 \right)^{g}$ is increasing with $g$ and $g \leq \frac{\beta}{2} + 1$. Since $3^{\beta/2 + 1} \leq 2^{\beta}$ whenever $\beta \geq 8$, we conclude that the result holds for $\beta \geq 8$. For the case where $\beta < 8$, the result can be verified through exhaustive checking.
						
			Next, we consider the case $g \geq \frac{\beta}{2}+1.$ Note that under this setup, since $t_0 \geq t_1 \geq \cdots \geq t_{g-1}$, it follows that $t_{g-1} =1$. Otherwise, if $t_{g-1} = 2$, we would have $\sum_{j=0}^{g-2} t_j \geq \beta$. 
			
			For this case, we prove the result by induction on $g$. The base case, corresponding to $g = \frac{\beta}{2}+1,$ follows from the previous paragraph. Therefore, assume that the result holds for all $g < \gamma$ and consider $g = \gamma > \frac{\beta}{2}+1$. Since $\gamma > \frac{\beta}{2}+1$, we have $t_{\gamma-2} = t_{\gamma-1} = 1$ which implies that $\sum_{j=0}^{\gamma-3} < \beta - 1$, since otherwise, if $\sum_{j=0}^{\gamma-3} = \beta -1$, then $\sum_{j=0}^{\gamma-2} t_j = \beta$ and we arrive at a contradiction. Thus, we have $\gamma-1 \geq \frac{\beta-1}{2} +1$ and so we can apply the inductive hypothesis to the first $\gamma-1$ components of the sequence $(n_0, \ldots, n_{g-1})$ and conclude that there are at most $2^{\beta-1}$ possible choices for the sequence $(n_0, n_1, \ldots, n_{\gamma-1})$. Since $t_{\gamma-1} = 1$, it follows that the total number of different options for the sequence $(n_0, n_1, \ldots, n_{\gamma-1})$ is at most $2 \cdot 2^{\beta-1}$. This completes the proof.
\end{IEEEproof}
		
		Recall the main idea behind the proof of Lemma~\ref{lem:ps1}, where we tacitly assumed that $g = \beta$. There, we used the binary representation of the integer $t$, where $t$ denotes the test outcome of the super-pool, to determine which of the tested subgroups involved infected individuals. In order to make this argument work, we formed the super-pool by adding $2^i$ samples from group $\cQ_i \subseteq \cG_i$ for $i \in [[\beta]]$, where $|\cQ_i| = \frac{|\cG_i|}{2}$. Next, the idea is to add $N_i$ samples from each group, where $N_i$ is chosen by considering a mixed-radix representation of the number $t$.	
		
		We say that $(b_0, b_1, \ldots, b_{g-1})$ is the $(t_0, t_1, \ldots, t_{g-1})$-mixed radix representation for $t$ if the following is true. Let $N_0 = 1$. For $i \in [g-1]$, let $N_i = (t_{i-1}+1) \cdot N_{i-1}.$ Note that when $t_0 = t_1 = t_2 = \cdots = t_{g-1} = 1$, $N_i = 2^{i}$. The mixed radix representation of $t$ is of the form $t = \sum_{i=0}^{g-1} b_i \cdot N_i$, where $b_i \leq t_i$. Note that under this setup since $b_i \leq t_i$, the sequence $(b_0, b_1, \ldots, b_{g-1}) \in [[t_0+1]] \times [[t_1+1]] \times \cdots \times [[t_{g-1}+1]]$ provides a unique representation and is invertible provided that $(t_0, t_1, \ldots, t_{g-1})$ is given. In other words, given the number $t$ we can uniquely determine the $i$-th digit in the $(t_0, t_1, \ldots, t_{g-1})$ mixed radix representation for $t$, which is $b_i$. Furthermore, as a result of Claim~\ref{cl:maxrep}, we know that $t \leq 2^{\beta}-1=m$.

		We are now ready to proceed with the proof. Suppose that $(\cG_0, \cG_1, \ldots, \cG_{g-1})$ is a $\beta$-minimal group. We will prove the result by induction and we will show the inductive step (since the base case follows from similar ideas).	
		
		For the inductive step, assume the statement holds when the group size is at most $2^{\alpha'}$ and consider the setup where the group size is $2^{\alpha} = 2^{\alpha'+1}$. Note that we have $(t_0, t_1, \ldots, t_{g-1})$-infected groups $\cG_{0}, \cG_1, \ldots, \cG_{g-1}$ each of size $2^{\alpha' + 1}$. We form our super-pool as follows. As before, for each $i \in [[g]]$, we select a subset $\cQ_i \subset \cG_i$ of size $|\cQ_i| = 2^{\alpha'}$. For each individual in $\cQ_i$ we add $N_i$ samples into the superpool, where $N_i$ is as defined in the previous paragraphs. 
		
		Let $t$ be the output of testing the resulting super-pool according to (\ref{eq:tret}) and let $b_i$ denote the $i$-th symbol of the $(t_0, t_1, \ldots, t_{g-1})$-mixed radix representation of $t$. 
		
		Note that based on $t$, we can determine the number of infected individuals in each of the subgroups  $\cQ_0,$ $\cG_0 \setminus \cQ_0,$ $\ldots,$ $\cQ_{g-1},$ $\cG_{g-1} \setminus \cQ_{g-1}$. In particular, since we know $t_i$, and given the output $b_i$ which can be recovered after testing the super-pool, we know that for all $i \in [[g]]$, the number of infected subjects in $\cQ_i$ is $b_i$ and the number of infected subjects in $\cG_i \setminus \cQ_i$ is $t_i - b_i$. Given this information, we can generate $\cG_0', \cG_1', \ldots, \cG'_{g'-1},$ where for $i \in [[g]],$ $\cG_i' \subseteq \cQ_i$ or $\cG_i' \subseteq \cG_i \setminus \cQ_i$, such that the collection is a $\beta$-minimal group. Thus, we can apply the inductive hypothesis to ${\bf{G}}$. This establishes that we can identify $\beta$ infected individuals using at most $\alpha$ tests and completes the proof.
	\end{IEEEproof}
	
\begin{figure}[h]
		\centering
		\includegraphics[width=17cm]{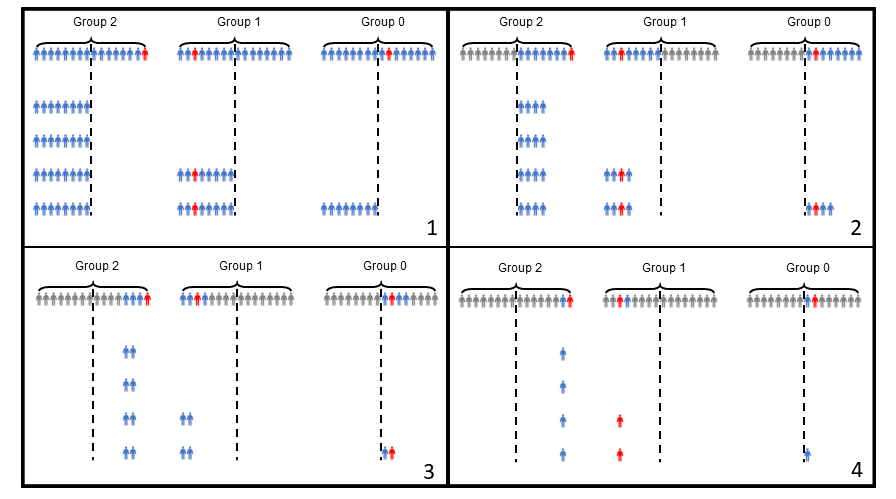}
		\caption{Intuitive illustration of the parallel search ASQGT procedure with some details removed for the ease of exposure. In this example, $m = 7$ and exactly one individual is infectious in each of the three groups. The weights of samples in each test round are set to $4,2,1$ as seen in Frame 1. A binary search procedure is implemented to find the infected individual in each group. In Frame $1$, the test outcome for the first round is $2,$ implying that there is one infected individual in the second group. Thus, the subgroups from groups $1$ and $3$ that were probed in Frame $1$ are discarded as illustrated in Frame $2$. Similarly, the second subgroup of group $2$ that was not tested is discarded as well. The subgroups that contain an infected individual are further probed as seen in Frames $2, 3$ and $4$. }\label{fig:1}
\end{figure}
	
\subsection{Deep search}\label{subsec:deep}
	
	Next, we consider the case $d < \beta$, and show that there exists an ASQGT scheme which requires roughly $\frac{d}{\beta - \log(\beta)} \cdot (\alpha + \log(\beta)) + d$ tests. Recall that the main idea behind the parallel search procedure was to simultaneously run a binary search on $g$ subpools each of size $2^{\alpha}$. In this manner, using $\alpha$ tests we can identify $\beta$ infected. For $d < \beta$, there are not sufficiently many infected individuals to use this method, and so for this setup, rather than perform a binary search in parallel, we test roughly $2^{\beta - \log(\beta)}$ (significantly smaller) subpools at the same time. We refer to this procedure as deep search. 
	
	Before proving a relevant lemma, we begin by describing a variant of  well-known Newton identities.  For completeness, we include a proof.
	
	\begin{claim}\label{cl:new} Let $\cS = \{{j_1, \ldots, j_d\}} \in \mathbb{Z}$ be a multiset of nonnegative integers each of which has value at most $p-1,$ where $p$ is an odd prime. Define $p_{\ell}(\cS) = \sum_{k=1}^d j_k^{\ell} \bmod p,$ the $\ell^{\text{th}}$ power sum of $\cS$ over the finite field $\mathbb{F}_p$. Then, one can recover $\cS$ given $\big (p_0(\cS), p_1(\cS), \ldots, p_{d}(\cS)\big)$.
	\end{claim}
	\begin{IEEEproof} We represent $\cS$ using $\cS^{(+)},$ containing the positive elements in $\cS$ and $z \in \mathbb{Z},$ which denotes the number of zeros in $\cS$. Given $\cS^{(+)}$ and $z$, the set $\cS$ is uniquely determined.
		
		First, note that Newton's identities can be used to recover the set 
		$\cS^{(+)}=\big\{ i_1, i_2, \ldots, i_{d'} \big\}$. To see this, let $\sigma(i_1, i_2, \ldots, i_{d'}) = \prod_{k=1}^{d'} (1-i_k x) = \sum_{k=0}^{d'} \sigma_k x^k \in \mathbb{F}_p[x]$ and assume that the operations are over the polynomial ring $\mathbb{F}_p[x],$ where the elements in $\cS$ are assumed to lie in $\mathbb{F}_p$. Then, we have
		\begin{align*}
		\sum_{\ell=1}^d p_{\ell}(\cS) \cdot x^\ell \, (\bmod x^{d+1}) &= \sum_{\ell=1}^d p_{\ell}(\cS^{(+)}) \cdot x^\ell \, (\bmod x^{d+1}) = 
		\sum_{\ell=1}^{d} \sum_{k=1}^{d'} i_k^{\ell} \cdot x^{\ell} \, (\bmod x^{d+1}) = \sum_{k=1}^{d'} \sum_{\ell=1}^{d}  i_k^{\ell} \cdot x^{\ell} \, (\bmod x^{d+1}) \\
		&=  \sum_{k=1}^{d'} \left( \frac{1-i_k^{d+1} \cdot x^{d+1}}{1 - i_k \cdot x} - 1 \right) \, (\bmod x^{d+1})= \sum_{k=1}^{d'}  \frac{i_k \cdot x}{1 - i_k \cdot x}  \, (\bmod x^{d+1}),
		\end{align*}
		which implies 
		$$\sum_{\ell=1}^d p_{\ell} \cdot (\cS^{(+)}) \cdot x^\ell  \cdot \sigma(i_1, \ldots, i_{d'}) \, (\bmod x^{d+1}) = - x \cdot \sigma'(i_1, \ldots, i_{d'}).$$
		The above equality in turn implies $\sum_{k=0}^{\ell-1} \sigma_k \cdot p_{\ell-k}(\cS)  = -\ell \cdot \sigma_\ell$. Thus, given $p_\ell(\cS),$ $\ell \in [d]$, one can recover $\sigma(\cS^{(+)})$ as well as the multiset $\cS^{(+)}$. The multiset $\cS$ can be subsequently recovered by noting that the number of zeros in $\cS$ equals $p_0(\cS) - |\cS^{(+)}|$.
	\end{IEEEproof}
	
	\begin{lemma}\label{lem:ds} Let $p$ be an odd prime such that $p \geq 2^L-1$ and $(p-1) \cdot d < 2^{\beta}$. Suppose that $\cG$ is a $d$-infected set of size $2^{\alpha}$, and $d \leq p-1$. Then we can identify the $d$ infected individuals using at most $ d \cdot \frac{\alpha}{L} $ tests.
	\end{lemma}
	\begin{IEEEproof} For simplicity we assume that $L | \alpha$, and, similar to Lemmas~\ref{lem:ps1} and \ref{lem:ps2}, use induction in $\alpha$. For the case $\alpha = L$, we run $d$ tests, and for each test we design a different test group. For $\ell \in [d]$, test group $\ell$ contains $j^{\ell} \in \mathbb{F}_{p}$ samples from each individual indexed by $j \in [[2^L]]$. Suppose that $\cD$ is a multi-set of elements from $[[2^L]]$ and that $\cD$ is such that if group $j$ has $k$ infected individuals, then the elements from group $j$ appear $k$ times in $\cD$. Then according to the above setup the output of performing the SQGT on pool $\ell$ results in the following $\ell$-th power sum:
		\begin{align*}
		p_{\ell}(j_1, j_2, \ldots, j_d) = \sum_{k=1}^d j^{\ell}_k.
		\end{align*}
		Note that $j_k^{\ell} < p$ (since by design $j^{\ell} \in \mathbb{F}_p$) and so $p_i(j_1, j_2, \ldots, j_d) \leq (p-1) d < 2^{\beta}$. Thus, for $\ell \in [[d+1]]$, we can recover $p_\ell(j_1, j_2, \ldots, j_d)=p_\ell(j_1, j_2, \ldots, j_d) \bmod p $, since $p_0(j_1, j_2, \ldots, j_d) \bmod p = d$ follows from the fact that $\cG$ is a $d$-infected set. From the set of $d+1$ power sums over the field $\mathbb{F}_p$, we can recover the multi-set $\{j_1, \ldots, j_d\}$ from Claim~\ref{cl:new}, which completes the proof of the base case.
		
		For the inductive step, assume the statement holds for group sizes at most $2^{\alpha'}$ and consider a group size $2^{\alpha} = 2^{\alpha'+L}$. As in the proofs of Lemmas~\ref{lem:ps1} and \ref{lem:ps2}, we work with subgroups. The subgroups are formed by partitioning the set of $2^{\alpha'+L}$ individuals into $2^L$ subgroups $\cP_1, \cP_2, \ldots, \cP_{2^L}$ each of size $2^{\alpha'}$. Applying the same ideas as before, we form $d$ test groups where test group $\ell \in [d]$ contains $j^\ell \in \mathbb{F}_p$ samples from each individual in subgroup $j \in [[2^L]]$. 
		
		Let $\cD=\{ j'_1, j'_2, \ldots, j'_d \}$ be a multiset of integers such that $j'_u$ appears $t$ times in the multiset if and only if group $j'_u$ has $t$ infected individuals. Using the same approach as for the base case, we first recover the power sums $p_i(j'_1, j'_2, \ldots, j'_d)$. Then from Claim~\ref{cl:new}, we recover the set $\cD$ in the same manner as before and we apply the inductive hypothesis to the subgroups in $\cD$. This completes the inductive step and the proof.
	\end{IEEEproof}
	
	\begin{remark} For the case $d=1$, the deep search procedure coincides with the approach described in Example~\ref{ex:motivate}. Deep search may be of limited practical value due to the large amounts of sample material required for testing, but is of theoretical relevance due to the fact that it generalizes Hwang's generalized splitting method to the SQGT setting for a small number of infected individuals.
	\end{remark}
	
	\subsection{$(n,d)$-ASQGT schemes}
	
	As discussed in the text following Example~\ref{ex:motivate}, our general approach to adaptive SQGT is to first partition the set of $n$ individuals into $\frac{d}{\beta}$ subpools and test each subpool separately using either parallel search or deep search, depending on the number of infected in each subpool. Parallel search produces the best results in the worst case, provided that the number of infected individuals across all the subpools is $\leq \beta,$ while parallel search gives the best results for the case of a large number of infected individuals.
	
	Let $T_{P} (n,d)$ denote the number of tests required by our ASQGT scheme, summarized in Algorithm~1, and let $n - d = 2^{\alpha} \cdot d + 2^{\alpha} \cdot \delta + \Delta$, where $\alpha, \delta, \Delta$ are integers such that $\delta < d$ and $\Delta < 2^{\alpha}$. In order to simplify the notation by avoiding floor and ceiling functions, we assume that $\beta | d$ and $\beta | \delta$. 
	
	\begin{theorem}\label{th:tgeq} $T_{P}(n,d) \leq \frac{d}{\beta} \cdot \left( \alpha + 3 + \log \beta\right) + \frac{\delta}{\beta}. $
	\end{theorem}
	\begin{IEEEproof} Since the first step involves testing $\frac{d}{\beta} + \frac{\delta}{\beta}$ groups, the first step requires $\frac{d}{\beta} + \frac{\delta}{\beta}$ tests. For the next steps, note that each group has size $\leq 2^{\alpha+1} \beta$. Hence, we can uncover $\beta$ infected individuals using at most $\frac{\alpha + 1 + \log(\beta)}{\beta}$ tests according to Lemmas~\ref{lem:ps1} and \ref{lem:ps2}. In step 3), we use one additional test for every $\beta$ infected individuals. Since there are $d$ infected the total number of tests required by Algorithm~1 equals 
		\begin{align*}
		T_{P} (n,d) \leq \left( \frac{d}{\beta} + \frac{\delta}{\beta} \right) + \frac{d}{\beta} \cdot (\alpha + 1 + \log(\beta) ) + \frac{d}{\beta} = \frac{d}{\beta} \cdot ( \alpha + 3 + \log(\beta) ) + \frac{\delta}{\beta}.
		\end{align*}
	\end{IEEEproof}
	
			\begin{algorithm}[ht]
		\caption{Parallel search ASQGT scheme}\label{alg:nd2}
		\begin{enumerate}
			\item \textit{\textbf{Initialize}}: Partition the set of $n$ individuals into $\frac{d+\delta}{\beta}$ groups, denoted by $\cG_0, \cG_1, \ldots, \cG_{\frac{d+\delta}{\beta}-1},$ each of size $\leq \beta \cdot 2^{\alpha+1}$. 
			
			Test each subgroup individually. For $i \in [[\frac{d+\delta}{\beta}]]$, suppose that $\cG_i$ is a $t_i$-infected group and let $D$ denote the total number of infected subjects across all groups.
			\item \textit{\textbf{Parallel Search}}: Identify a $\beta$-minimal group $(\cG_{i_0}, \ldots, \cG_{i_{g-1}})$, and apply parallel search on the group to uncover $\beta$ infected individuals. Remove the $\beta$ infected individuals from their respective groups. 
			\item \textit{\textbf{Update}}: Use one additional test to determine the number of infected subjects in $(\cG_{i_0}, \ldots, \cG_{i_{g-1}})$ after Step 2). Update $t_{i_0}, \ldots, t_{i_{g-1}}$ and $D$. If $D>0$, go to Step 2).
		\end{enumerate}
	\end{algorithm}
	
	As discussed earlier, the parallel search ASQGT scheme requires $\cO(\frac{d}{\beta})$ more tests than the information-theoretic lower bound. When $\beta=1$, our scheme requires $\cO(d)$ additional tests which agrees with the traditional adaptive binary setting studied in~\cite{H72}. 
	
	Next, we consider the second approach to the ASQGT problem based on deep search, for the case where $d < \beta$. Let $T_{D} (n,d)$ denote the number of tests required by our algorithm and, with a slight abuse of parameter definitions, assume that $n - d = 2^{\alpha} \cdot d$. Furthermore, assume as before that $d | 2^{\beta}$ and $d | n$. The corresponding approach is described in Algorithm~\ref{alg:nd}.

	\begin{algorithm}[ht]
		\caption{Deep search ASQGT scheme}\label{alg:nd}
		\begin{enumerate}
			\item \textit{\textbf{Initialize}}: Partition the set of $n$ individuals into $d$ groups, denoted by $\cG_0, \cG_1, \ldots, \cG_{d-1},$ each of size $2^{\alpha}$. Test each subgroup individually and let $D$ denote the total number of infected subjects across all groups.
			\item \textit{\textbf{Deep Search}}: Identify a $t_i$-infected group $\cG_i$, and apply deep search to uncover $t_i$ infected subjects, for some $i \in [[d]]$. 
			\item \textit{\textbf{Update}}: Let $D = D - t_i$. If $D > 0$, go to Step 2). 
		\end{enumerate}
	\end{algorithm}
	
	\begin{theorem}\label{th:tgeq2} 
	The number of tests for deep search ASQGT satisfies
	$$T_{D}(n,d) \leq d \cdot \frac{\alpha}{\beta - \log \beta-1} + \beta.$$
	\end{theorem}
	\begin{IEEEproof} The first step in Algorithm~2 requires $d < \beta$ tests. According to Lemma~\ref{lem:ds}, Step 2) requires at most $t_i \cdot \frac{\alpha}{\beta - \log(\beta)-1}$ tests. Hence, the total number of tests is upper bounded as
		\begin{align*}
		T_{D}(n,d) \leq \beta + \sum_{i=0}^{d-1} t_i \cdot  \frac{\alpha}{\beta - \log \beta-1} = \beta + d \cdot \frac{\alpha}{\beta - \log \beta -1}.
		\end{align*}
	\end{IEEEproof}	

\subsection{Error-resilient $(n,d)$-ASQGT schemes}
	
	We consider next the question of designing ASQGT models that can tolerate a bounded number of birth-death (BD) chain errors. Recall from (\ref{eq:tret}) that in the event that there are no errors, the output of testing a pool of individuals, of which $d$ are infected, is an integer $t$, such that $t = d$ for $d \leq m$ and $t=m,$ whenever $d > m$. Suppose instead that the erroneous output of testing a pool is $t'$, where $t' \in \{t-1, t+1\}$ with the appropriate boundary conditions. We refer to such an error as a single BD error.
	
	Our main result is described in Theorem~\ref{th:noisy}. We prove that there exists a scheme that requires $\frac{d}{\beta-2} \cdot \left( \alpha + 3 + \log \beta \right) + \frac{\delta}{\beta}$ tests that can correct an arbitrary number of test errors. For the case where the number of test errors is a small integer $e$, $\left( \frac{d}{\beta} + e \right) \cdot \left( \alpha + 3 + \log \beta \right) + 2 \cdot \frac{d}{\beta} + \frac{\delta}{\beta} + e$ tests suffice, which implies that only $e \left( \alpha + 3 + \log \beta \right) + 2 \frac{d}{\beta} + e$ additional tests are required to correct $e$ errors in Algorithm~1.

The next claim highlights one of the main ideas behind our approach: Take multiple copies of samples from each of the individuals being tested in such a way to get error-free readouts even when errors occur. Here, as before, we assume that $2^\beta = m+1$.

\begin{claim}\label{cl:corr} Let $\cP$ be a pool of individuals and suppose that $\cP^{(\times 3)}$ is a pool which contains three samples from each individual in $\cP$. Let $t$ be the output of the test performed on the pool $\cP^{(\times 3)}$ given that no errors occur, and suppose $t'$ is a possibly erroneous output of the test performed on the pool $\cP^{(\times 3)}$. Given $t'$, one can determine $t$.   \end{claim}
\begin{IEEEproof} Since we have taken $3$ samples from each of the individuals in the pool $\cP$, it follows that $t \, (\bmod 3) =0$. Thus, if an error occurs, the output of the test under the BD model equals $t' \in \{ t+1, t-1 \}$ which implies that $t' \, (\bmod 3) = \pm 1$. If $t' \bmod 3 = 1$, then $t' = t+1$ and so we can recover $t$ by simply decrementing $t'$ by one. Similarly, if $t' \bmod 3 = -1$, then $t'=t-1$ and we can recover $t$ by incrementing $t'$ by one.
\end{IEEEproof}

Using the idea from the previous claim, we can determine exactly how many infected individuals are present in each of the tested pools despite the fact that testing errors can occur. We describe the underlying method through an example, for which we need the following terminology.

We say that \textit{\textbf{$\cP_i$ is a $t_i$-infected group}} if the output of testing $\cP^{(\times 3)}_i$ is in the set $\{ 3 t_i-1, 3t_i, 3t_i + 1 \}$. We also say that \textit{\textbf{$(\cP_{i_0}, \ldots, \cP_{i_{g-1}})$ is a $\beta$-minimal group}} if $t_0 \geq t_{i_1} \geq \cdots \geq t_{i_{g-1}}$,  $\sum_{j=0}^{g-2} t_j < \beta$, but $\sum_{j=0}^{g-1} t_j \geq \beta.$

\begin{example}\label{ex:robust} For simplicity, assume that we have $m = 3\left(2^{\gamma} - 1\right)$ thresholds, and suppose that $(\cP_0, \cP_1, \ldots, \cP_{g-1)}$ is a $\gamma$-minimal group, where we again make a simplifying assumption, namely $\gamma = g$. We proceed in the same manner as described in Lemma~6 and we first form a super-pool, denoted $\overline{\cP}$ which consists of $2^i$ copies of each sample in $\cP_i$. Afterward, we generate a larger pool of samples, $\overline{\cP^{(\times 3)}}$ which contains $3$ copies of each sample in $\overline{\cP}$. 

Notice that given the output of the test $\overline{\cP^{(3)}}$, we can uniquely determine the number of infected that are in each of the groups $\cP_0, \cP_1, \ldots, \cP_{g-1}$. Suppose that $t'$ is the output of testing $\overline{\cP^{(3)}}$ and suppose $t$ is the output of testing $\overline{\cP^{(3)}}$ assuming no errors occur during testing. From Claim~\ref{cl:corr}, we can recover $t$ and from Lemma 6 it is possible to determine how many infected are in $\cP_0$, how many infected are in $\cP_1$, etc.
	
		Another simple way to see how the above scheme overcomes BD noise is to see that it suffices that test outcomes differ from one another by at least three. This can be easily accomplished by \emph{fixing} the coefficients of $2^1$ and $2^0$ in the binary representation of the test outcomes to zero.
		
		More precisely, we can artificially introduce two subgroups, so that when $m = 2^{\gamma} - 1$, we collect samples from subgroups labeled by $ 1 < i < \gamma$, $2^{i}$ with the amounts dictated by their labels. If the observed test outcome is $t' = \sum_{i =0 }^{\gamma-1} \bar{e}_i \cdot 2^i$, then the true test outcome is decoded as:
		\begin{align}
		e_{\gamma -1} e_{\gamma -2} \dots e_ 2 e_1 e_0 =
		\begin{cases}
		\bar{e}_{\gamma -1} \bar{e}_{\gamma -2} \dots \bar{e}_2 00, &\text{ if } \bar{e}_1 \bar{e}_0 = 00 \text{ or } \bar{e}_1 \bar{e}_0 = 01 ,\\
		\bar{e}_{\gamma -1} \bar{e}_{\gamma -2} \dots \bar{e}_2 \bar{e}_1 \bar{e}_0 +1 \text{ (binary addition)}, &\text{ if } \bar{e}_1 \bar{e}_0 = 11.
		\end{cases}
		\end{align}
\end{example}

The following claim is straightforward.

\begin{claim}\label{cl:robust} Let $\alpha, \beta \geq 2,g$ be positive integers where $2^\beta = m+1$, $g \leq \beta-2$. Suppose that $(\cG_0, \cG_1, \ldots, \cG_{g-1})$ is a $(\beta-2)$-minimal group and that each group has size at most $2^{\alpha}$. Then we can identify $\beta-2$ infected individuals using at most $\alpha$ tests.
\end{claim}
\begin{IEEEproof} The proof follows immediately by applying the procedure described in Example~\ref{ex:robust} and noting that $3 \cdot 2^{\beta-2} < 2^{\beta}  -1$ when $\beta \geq 2$.
\end{IEEEproof}

Next, we turn our attention to a scheme designed for a small number of testing errors $e$. To this end, let $T_{N} (n,d,e)$ denote the number of tests required for a noisy ASQGT scheme that tolerates up to $e$ BD testing errors (see Algorithm~3). As before, let $n - d = 2^{\alpha} d + 2^{\alpha} \delta + \Delta$, where $\alpha, \delta$ and $\Delta$ are integers such that $\delta < d$ and $\Delta < 2^{\alpha}$. Once again we assume that $\beta | d$ and $\beta | \delta$.

 \begin{algorithm}[ht]
		\caption{Noisy search ASQGT scheme}\label{alg:nd2noisy}
		\begin{enumerate}
			\item \textit{\textbf{Initialize}}: Partition the set of samples from the $n$ individuals into $\frac{d+\delta}{\beta}$ groups, denoted by $\cP_0, \cP_1, \ldots, \cP_{\frac{d+\delta}{\beta}-1}$, and each of size at most $\beta 2^{\alpha+1}$. 
			
			Test each subgroup $P_i^{(\times 3)}$ individually. For $i \in [[\frac{d+\delta}{\beta}]]$, suppose that $\cP_i$ is a $t_i$-infected group and let $D$ denote the total number of infected subjects in all subgroups.
			\item \textit{\textbf{Parallel Search}}: Identify a $\beta$-minimal group $(\cP_{i_0}, \ldots, \cP_{i_{g-1}})$, and apply parallel search to uncover $\beta$ potential infected subjects. Divide the set of $\beta$ potentially infected individuals into two groups of sizes $\lfloor \frac{\beta}{2} \rfloor$ and $\lceil \frac{\beta}{2} \rceil$, denoted by $\cD^{(\times 3)}_1, \cD^{(\times 3)}_2$.
			\item \textit{\textbf{Verify}}: Test $\cD^{(\times 3)}_1, \cD^{(\times 3)}_2$ to determine the total number of infected recovered. Update $t_{i_0}, \ldots, t_{i_{g-1}}$ and $D$. 
			\item \textit{\textbf{Update Large Group Counts}}: If only one group is present, $|\cP_{i_0}| \geq \beta$, and $t_0 \geq 1$, then test $\cP_{i_0}^{(\times 3)}$ to determine the number of infected in $\cP_{i_0}$. Go back to Step 2).
			\end{enumerate}
	\end{algorithm}
	
	We prove the correctness of our algorithm in the following theorem.
	
	\begin{theorem}\label{th:noisy} Let $\beta \geq 2$. We have
	$$ T_N(n,d,e) \leq \min \left( \left( \frac{d}{\beta} + e \right) \cdot \left( \alpha + 3 + \log \beta \right) + 2 \cdot \frac{d}{\beta} + \frac{\delta}{\beta} + e, \; \frac{d}{\beta-2} \cdot \left( \alpha + 3 + \log \beta \right) + \frac{\delta}{\beta} \right).$$
	\end{theorem}
	\begin{IEEEproof} The second term under the minimum follows immediately from the parallel search ASQGT scheme given the use of a robust parallel search. Therefore, in the remainder of the proof, we focus our attention on the first term.
	
	The first step of our algorithm requires $\frac{d+\delta}{\beta}$ tests and each time we execute Step 2), we perform $\alpha + 1 + \log \beta$ tests. Since Step 2) is executed at most $\frac{d}{\beta} + e$ times this implies that the total number of tests required by the first two steps of our procedure is at most 
	$$ \frac{d + \delta}{\beta} + \left( \frac{d}{\beta} + e \right) \cdot \left( \alpha + 1+ \log \beta \right).$$
	
	For Step 3), note that since $\max \Big \{ |\cD_1^{(\times 3)}|, |\cD^{(\times 3)}_2| \Big \} \leq \lceil \frac{\beta}{2} \rceil$ we have $t' \leq 3 (2^{\lceil \frac{\beta}{2} \rceil} -1) < 2^{\beta}-1 = m$ when $\beta \geq 2$, and so we can determine exactly how many infected subjects are in each of the sets $\cD^{(\times 3)}_1, \cD^{(\times 3)}_2$ in  Step 3). Each time Step 3) is executed, we require $2$ tests. Since Step 2) is executed at most $\frac{d}{\beta} + e$ times, this step requires at most
	$$ \frac{2 \cdot d}{\beta} + 2 \cdot e$$
	tests.
	Finally, since Step 4) is executed at most $\frac{d}{\beta} + e$ times, it follows that the total number of tests is at most $\frac{d+\delta}{\beta} + \left( \frac{d}{\beta} + e \right) \cdot \left( \alpha + 1 + \log \beta \right) + \frac{2 \cdot d}{\beta} + 2 \cdot e + \frac{d}{\beta} + e$, which proves the claimed result.
	\end{IEEEproof}

We conclude the above exposition by observing that in a very recent companion paper~\cite{cheraghchi2021semiquantitative}, we described adaptive schemes for SQGT that only use two rounds of testing and may hence have practical advantages over deep search methods. Nevertheless, the results in~\cite{cheraghchi2021semiquantitative} rely on nonconstructive expander graph existence guarantees and trade other desirable testing properties for a reduced number of testing rounds.

	\subsection{Extensions to nonuniform threshold widths}
		
	The next two examples illustrate how the ideas from the previous sections can be extended to the case where the threshold widths increase exponentially. For this case, we only consider small values of $m$ (i.e., $m=3,4$). 
	
	\begin{example}\label{ex:widths} In the following, we consider a model that mirrors the results from Section~\ref{twostage}. Suppose that the test outcomes equal
		\begin{align}\label{eq:tret22}
		t=\begin{cases}
		0, &\text{ if there are no infected subjects in the test},\\
		1, &\text{ if the number of infected samples is $1$}, \\
		2, &\text{ if the number of infected samples is in $[2,3]$}, \\
		3 &\text{ if the number of infected samples is in $[4,7]$}.
		\end{cases}
		\end{align}
		
		We consider the following extension of the approach discussed in Example~\ref{ex:motivate}. Suppose we have a pool of size $2^{\alpha}$ that contains at least one infected subject. We start by testing this pool to determine the total number of infected individuals. There are two cases to consider: (a) The output of the test is $2$ or $3,$ which indicates that there is more than a single infected in the pool or (b) The output of the test is $1$. 
		
		Suppose that the outcome is (b). In this case, we run a variant of deep search. In particular, we divide the pool into $4$ subpools and form a superpool from these $4$ subpools which contains $0$ samples from the first pool, $1$ sample from the second pool, $2$ samples from the third pool, and $4$ samples from the fourth pool. It is straightforward to verify that in this case we can determine which of the  $4$ subpools contain the single infected sample by testing the superpool, and we then repeat this procedure using the subpool which contains the single infected.
		
		If the outcome is (a), then we proceed to divide the pool of size $2^{\alpha}$ into two disjoint subpools, each of size $2^{\alpha-1}$. We further select one of the two subpools for testing. If the subpool tested contains a single infected, then we continue by applying the procedure discussed in the previous paragraph on the subpool of size $2^{\alpha-1}$ that contains one infected sample. Otherwise, we repeat the procedure from this paragraph on one of the subpools of size $2^{\alpha-1}$ that contains more than a single infected subject.
		
		Using the procedure described above, it is straightforward to verify that recovering $2$ infected individuals requires at most $\alpha$ tests provided we know the number of infected samples in the pool of size $2^{\alpha}$. Now suppose $n = d \, 2^{\alpha}$. Then we can recover $d$ infected subjects using at most $2 \cdot d + \frac{d \cdot \alpha}{2}$ tests as follows. First, we partition the set of $n$ individuals into $d$ groups each of size $2^{\alpha}$ and we initially test each of these $d$ groups. Afterward, we search for the infected individuals using the process outlined in this example.
	\end{example}
	
	We note that despite the fact that we have focused on the case where $m$ is a power of two, the next example shows that in some cases our ideas extend to settings where $m$ is not necessarily a power of two. In the next example, we show an adaptive scheme that requires at most roughly $d + d \cdot \left( \log_3(\frac{n}{d}) + 1 \right)$. The ideas are similar to the previous example except that here we only allow $3$ thresholds.
	
	\begin{example} For this example, we assume that $n = d \cdot 3^{\alpha}$. The output of the test is $t$, where:
		\begin{align}\label{eq:tret2}
		t=\begin{cases}
		0, &\text{ if no infected samples are present in the pool},\\
		1, &\text{ if the number of infected samples equals $1$}, \\
		2, &\text{ if the number of infected samples is $>1$.} \\
		\end{cases}
		\end{align}
		The core idea behind the testing strategy is a simple extension of the previous example. Suppose we have a pool of size $3^{\alpha}$ that contains at least one infected individual. First, we test this pool of size $3^{\alpha}$ to determine the total number of infected individuals. There are two cases to consider: (a) The output of the test equals $2,$ which indicates that there is more than one infected sample in the pool or (b) The output of the test equals $1$. 
		
		Suppose (b) occurred. We perform the same procedure as before except that instead of dividing the pool into $4$ subpools, we divide the pool into $3$ subpools of equal size. Next, we form a superpool from these $3$ sub pools which contains $0$ samples from the first pool, $1$ sample from the second pool, and $2$ samples from the third pool. Similarly as before, we can determine which of the three subpools contains the single infected sample, and we then repeat this procedure using the subpool which contains the single infected sample.
		
		If (a) occurs, then we perform the same procedure as in the previous example. In particular, we divide the pool of size $3^{\alpha}$ into two disjoint subpools each of size at most $\lceil \frac{3^{\alpha}}{2} \rceil$ and perform a single test. If two infected individuals are contained in a single pool, then we repeat the procedure from this paragraph on the pool of size at most $\lceil \frac{3^{\alpha}}{2} \rceil$ that contains at least two infected samples. Otherwise, we perform the procedure from the previous paragraph on the subpool of size at most $\lceil \frac{3^{\alpha}}{2} \rceil$ that contains a single infected sample.
		
		Using this approach, it is straightforward to verify that recovering an infected requires at most $\alpha + 1$ tests. Thus we can recover $d$ infected individuals using at most $d + d \cdot \left( \alpha + 1\right)$ tests as follows. First, we partition the set of $n$ infected into $d$ groups each of size $3^{\alpha}$. Afterward, we search for the infected individuals using the process outlined in this example.
	\end{example}
	
\begin{figure}[h!]
		\centering
		\includegraphics[scale=.55]{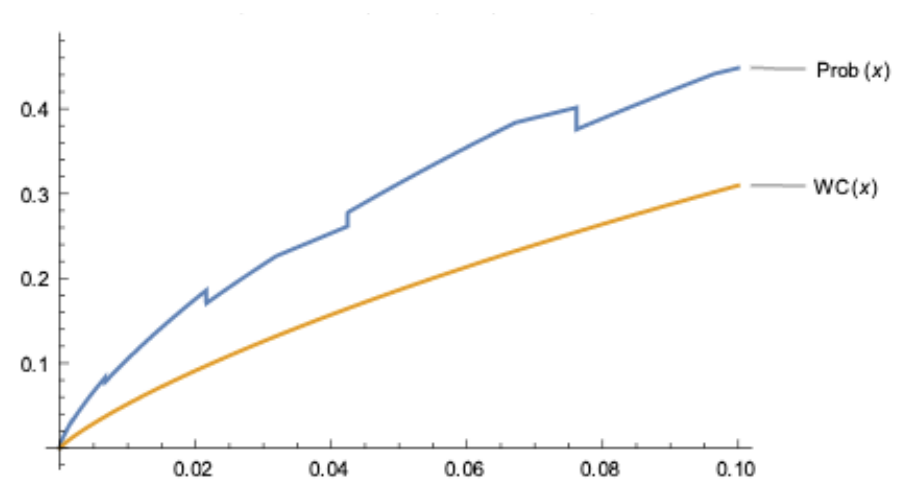}
		\caption{Comparison of the average number of test per individual for the $3$-threshold schemes (\ref{eq:jr}) and  (\ref{eq:rg}). The choice of $s$ is optimized for each value of $p$ for the probabilistic priors setting.}\label{fig:compjr}
	\end{figure}

	Since the model proposed in the previous example is identical to the one used for probabilistic priors and described in Section~\ref{twostage}, we now directly compare the two in terms of the number of tests required per individual. Recall that the model in section~\ref{twostage} required on average
	\begin{align}\label{eq:jr}
	\frac{1}{s} + p \cdot (1-p)^{s-1} \cdot \lceil \log s \rceil + 1 - (1-p)^s - s \cdot  p \cdot  (1-p)^{s-1}
	\end{align} 
	tests per individual where $s$ represents the size of each subpool used in the first step of the corresponding algorithm. For $n$ large enough, our setup requires approximately 
	\begin{align}\label{eq:rg}
	p + p \cdot \log_3 \left(\frac{1}{p} \right)
	\end{align}
	tests where $ p = \frac{d}{n}$. Figure~\ref{fig:compjr} compares the number of tests required in  (\ref{eq:jr}) and (\ref{eq:rg}).
	Notice that the ASQGT scheme from the previous example requires only roughly half the tests per individual of the approach from Section~\ref{twostage}. However, the latter method is simpler to implement in practice since it only requires two stages of testing. We also note that since the schemes are not exclusive, it is possible to use a combination of both approaches if needed.
		
	\section{A short note on community-aware testing} \label{ongoing}
	
	As mentioned previously, in order to formulate effective and optimal testing schemes, the underlying community network structure must be incorporated. Towards that end, we assume that the community labels are known as well as their sizes (not the entire network, but entities like families or clusters of families in close proximity; this assumption is realistic, as most testing sites require subjects to submit their addresses). The aim is to find efficient strategies that will identify \emph{communities with high infection rates}, rather than infected individuals, as this would guide effective quarantine strategies. 
	
	More precisely, consider a partition of a population of $N$ individuals into communities $\cA_1,\cA_2,\dots,\cA_f$ of size $N_1, N_2, ..., N_f$, respectively, where $N=N_1+...+N_f$ and $f \geq 2$. Each group has some (unknown) number of infected individuals equal to $d_i$, $i=1,\ldots,s$, and $d=\sum_{i=1}^f d_i$. The following question is of interest: Devise adaptive and nonadaptive GT and SQGT schemes that identify heavy hitter communities, i.e., communities with at least $d/k$ infected individuals, where $k$ is an input parameter. 
	
	The naive nonadaptive scheme for this problem corresponds to running within each of the $f$ communities an optimal testing scheme for determining the number of infected individuals. Each such scheme requires $\Theta(\log N_i)$ tests~\cite{DM10}, leading to a total of $\Theta(\sum_{i=1}^f \log N_i)$ tests. Note that $\Theta(\log N)$ tests are both necessary and sufficient to estimate the number of defectives in general, but it is conceivable that a better result is possible when we already know an upper bound on this number, which is the case here. 
	This approach is suboptimal, and we describe an alternative nonadaptive binary testing scheme for the heavy hitters problem that requires significantly fewer tests when the total number of infected individuals, $d$, is much smaller than the population size $n$.
	We leave as an important open problem to construct semiquantitative testing schemes for the heavy hitters problem with few adaptivity stages and requiring significantly fewer tests than the scheme we propose below.
	
	First, we note that the heavy hitters problem can be solved with $t=O(k\log f)$ queries that on input a set $T\subseteq [f]$ output $\sum_{i\in T}d_i$~\cite{indyk2013sketching}, which corresponds to the quantitive GT model. Indeed, this corresponds to the setting of compressed sensing with 0/1 linear tests.
	We show below how to emulate such a query with error probability $\eps$ using $r=O(2^{2d}\cdot \log(1/\eps))$ randomized disjunctive queries.
	Combining the two observations above immediately yields a nonadaptive group testing scheme using $r \cdot t=O(k \cdot \log f \cdot 2^{2d}\cdot \log(1/\eps))$ tests that solves the heavy hitters problem with error probability at most $t\cdot\eps$ via a union bound.
	For example, if $d<\log\log\log N$ and $\eps\leq \frac{1}{t\log n}$ then the required number of tests may be significantly smaller than that required by the naive scheme from the previous paragraph.
	
	It remains to show how to emulate the query $\sum_{i\in T}d_i$ with error probability $\eps$ using $r$ tests.
	Consider a randomized test obtained by independently including each individual $j\in\bigcup_{i\in T}\cA_i$ in the test with probability $1/2$, and let $Y$ denote the test output.
	Then, we have $\Pr[Y=0]=\prod_{i\in T}2^{-d_i}=2^{-\sum_{i\in T}d_i}$, since the test outputs $0$ if and only if no infected individual is included.
	Noting that $\sum_{i\in T}d_i\leq d$, a Chernoff bound guarantees that we can determine $\sum_{i\in T}d_i$ with error probability at most $\eps$ by independently sampling $r=O(2^{2 d} \log(1/\eps))$ such tests $Y_1,Y_2,\dots,Y_r$ and setting our estimate to be $\left\lceil -\log\left(\frac{1}{r}\sum_{a=1}^r Y_a\right)\right\rfloor$, where $\lceil x\rfloor$ denotes the closest integer to $x$.
	This yields the desired result.

We conclude the discussion with the following remarks:
\begin{enumerate}
\item The above scheme provides a reduction from 
community-aware testing of heavy hitters to the standard
heavy hitters problem in compressed sensing. While this 
provides a proof of concept, it remains to be seen whether a
direct approach can provide more effective test designs for
the identification of heavy hitter communities via
disjunctive queries.

\item An important direction is to construct an analogous 
approach for the detection of heavy hitter communities
using SQGT schemes. Recall that quantitative group testing
(equivalently, the compressed sensing model via binary matrices)
is an extremal special case of SQGT, whereas standard group
testing is another extreme. It is therefore natural to expect
that SQGT provides the identification of heavy hitter communities
at varying efficiency depending on the granularity of the 
quantization levels.

\item Furthermore, we ask whether adaptivity can help
to identify heavy hitter communities more efficiently than
nonadaptive schemes can offer.
\end{enumerate}

\section{Mutations and RT-PCR noise} \label{modeling}	
A number of works have focused on RT-PCR asymmetric error models that assume that positive samples can actually test negative while the opposite scenario is highly unlikely~\cite{nikolopoulos2020community}. As already pointed out, these assumptions are not practically justified since even as few as $10$ viral cDNA fragments can lead to detectable fluorescence levels after roughly $40$ cycles of amplification. Hence, the cause of false negative measurements does not lie in inadequate PCR testing but erroneous sample collection instead, in addition to possible mutations in the genomic regions used as primers. In both cases, no matter how many times an RT-PCR test is repeated, an infected sample may not be identified (i.e., the sample is masked). As an example, the CDC originally identified three pairs of primers from the N open reading frame (gene) of the SARS-CoV-2 virus for testing, but since one pair was removed due to the presence of mutations in a larger-than-acceptable population. There are further efforts to reduce the problem of false negatives due to mutations such as using primers from at least two genes~\cite{park2020optimization}. 

The problem of mutations in primer regions of individuals tested via RT-PCR is of relevance to GT both from the perspective of measurement modeling (as mutations add an additional level of nonlinearity to the measurements that are not captured by current approaches) as well as error analysis. It is important to observe that mutations or undesired hybridization to nontarget regions may lead to variable test outcomes for the same sample in different tests. As a result, trying to exactly estimate the viral load of the individuals as proposed in~\cite{tapestry} is not possible and estimates like the ones used in the SQGT framework may be more appropriate. Furthermore, quantization noise is signal-dependent, which is another desirable feature of the SQGT framework.  	
 
To examine the influence of mutations on RT-PCR we examined the GISAID~\cite{gisaid} database of Cov-SARS-2 genomes and identified $7$ individuals with mutations in the N1 and N2 primer regions. Using the \textit{FastPCR online} simulation software~\cite{primerdigital}, we examined the influence of the mutation on primer binding and PCR amplification. The results are summarized in Table~\ref{tab:my-table}. Furthermore, the complete primer and DNA sequences are given in Appendix~\ref{app1}. There, the symbols `f' and `r' refer to the DNA strands' forward and reverse directions, respectively. In the forward direction, the genome and primer have to be an exact match, while in the reverse direction the two strings have to be Watson-Crick complementary. As one can observe, mutations along the forward (reverse) direction of the N1 or N2 regions that do not contain the exact match (or Watson-Crick complement) can severely affect the efficiency of primer/target bonding. For a more precise characterization, the corresponding melting temperatures are given in the Table~\ref{tab:my-table} along with an estimate of the primer binding efficiency for the N1 and N2 regions. 	
	
	\begin{table}[]
\centering
		\caption{Simulation results}
		\label{tab:my-table}
		\begin{tabular}{|l|l|l|l|l|}
			\hline
			Patient                         & Primer Region & Amplification Predicted? & Melting Temperature (in $^\degree$ C) & Amplification \% \\ \hline
			\multirow{4}{*}{EPI ISL 413609} & N1f           & Y                        & 51.6                             & 100              \\ \cline{2-5} 
			& N1r           & Y                        & 56.3                             & 100              \\ \cline{2-5} 
			& N2f           & Y                        & 52.9                             & 100              \\ \cline{2-5} 
			& N2r           & Y                        & 52.2                             & 97               \\ \hline
			\multirow{4}{*}{EPI ISL 415600} & N1f           & Y                        & 51.6                             & 100              \\ \cline{2-5} 
			& N1r           & N                        & 51.7                             & 97               \\ \cline{2-5} 
			& N2f           & Y                        & 52.9                             & 100              \\ \cline{2-5} 
			& N2r           & Y                        & 54.7                             & 100              \\ \hline
			\multirow{4}{*}{EPI ISL 416650} & N1f           & N                        & 42.6                             & 97               \\ \cline{2-5} 
			& N1r           & Y                        & 56.3                             & 100              \\ \cline{2-5} 
			& N2f           & Y                        & 52.9                             & 100              \\ \cline{2-5} 
			& N2r           & Y                        & 54.7                             & 100              \\ \hline
			\multirow{4}{*}{EPI ISL 417938} & N1f           & Y                        & 51.6                             & 100              \\ \cline{2-5} 
			& N1r           & Y                        & 56.3                             & 100              \\ \cline{2-5} 
			& N2f           & Y                        & 47.3                             & 95               \\ \cline{2-5} 
			& N2r           & Y                        & 54.7                             & 100              \\ \hline
			\multirow{4}{*}{EPI ISL 422983} & N1f           & Y                        & 44.6                             & 95               \\ \cline{2-5} 
			& N1r           & N                        & 52.1                             & 95               \\ \cline{2-5} 
			& N2f           & Y                        & 52.9                             & 100              \\ \cline{2-5} 
			& N2r           & Y                        & 54.7                             & 100              \\ \hline
			\multirow{4}{*}{EPI ISL 424955} & N1f           & Y                        & 51.6                             & 100              \\ \cline{2-5} 
			& N1r           & Y                        & 56.3                             & 100              \\ \cline{2-5} 
			& N2f           & N                        & 43.1                             & 97               \\ \cline{2-5} 
			& N2r           & Y                        & 54.7                             & 100              \\ \hline
			\multirow{4}{*}{EPI ISL 425148}  & N1f           & Y                        & 51.6                             & 100              \\ \cline{2-5} 
			& N1r           & N                        & 51.7                             & 97               \\ \cline{2-5} 
			& N2f           & Y                        & 52.9                             & 100              \\ \cline{2-5} 
			& N2r           & Y                        & 54.7                             & 100              \\ \hline
		\end{tabular}
	\end{table}	
	As seen in the results of the simulation, not all samples that have mutations along the primer region are amplified. 	
	
\section{Conclusions} \label{sec:conclude}

We provided an in-depth description of the quantitative RT-PCR protocol suitable for nonexperts, an overview of existing GT testing protocols for Covid-19 and their practical implications. These comparative studies motivated further explorations of quantized GT (or semiquantitative GT (SQGT)) protocols, especially under a new measurement-error model termed the birth-death chain noise model. We furthermore developed state-of-the-art adaptive SQGT schemes with probabilistic and combinatorial priors and provided extensive analytical results, including performance bounds, algorithmic solutions, and noisy testing protocols. We also designed a probabilistic setting protocol which is extremely simple to implement by nonexperts and capable to handle heavy hitters. 

Many open problems remain, including:
\begin{itemize}
\item \textit{Probabilistic testing schemes for more than $3$ thresholds}: In Section~\ref{twostage}, we considered the setup where each test generated the output $0,1,$ or $2$ depending upon the number of defectives in each group. How much can one reduce the number of tests of our schemes if we incorporate additional semi-quantitative information, in the presence of errors?

\item \textit{Worst-case general SQGT testing schemes with a constant number of rounds}: The schemes described in Section~\ref{sec:worst} have the potential drawback that almost every test depends upon the results of prior tests. It has been shown that in the binary group testing setting, the information-theoretic lower bound can be achieved using only two rounds of nonadaptive testing when the number of infected individuals is at most $n^c$ for any constant $c<1$~\cite{HL19}. In a very recent line of work, we showed how to implement two-round SQGT schemes for the saturation model studied in Section~\ref{sec:worst}. It remains an open problem to generalize the approach for general quantized GT paradigms.

\item \textit{Practical SQGT schemes resilient to errors:} Our practical two-stage SQGT schemes from Section~\ref{sec:practical} can be enhanced with noise-resilience properties in a straightforward by repeating each test a prescribed number of times, while keeping the number of testing stages the same. Nevertheless, it would be interesting to find more efficient, and still practical, ways of adding good noise-resilience properties to these schemes.

\item \textit{Community-Aware Testing}: Section~\ref{ongoing} presented a simple approach to the problem of heavy-hitter community identification for classical GT. No results on this problem are currently available for general SQGT methods.
\end{itemize}

\appendices
	\section{Simulation results: PCR on DNA strands with mutations along primer regions} \label{app1}
	
	We present below the results of the PCR simulation run on DNA sequences that contain mutations along the N1 and N2 regions of the genome. The notation EPI ISL xxxxxx corresponds to the sample ID. As already indicated, the symbols `f' and 'r' at the end of the primer regions indicate the DNA strand directions, forward and reverse respectively. The symbol `Y' indicates a successful PCR amplification, while the symbol `N' indicates that PCR amplification cannot be initiated. We also list the percentage of the primer string that is matched by genomic DNA, and  the melting temperature $T_m$.  \\
	
	EPI ISL 413609
	\\
	N1f
	\zz 5-gaccccaaaatcagcgaaat !	Y 100\% $T_m$=51.6$^{\circ}$C
	\zz..||||||||||||||||||||	!
	\zz tggaccccaaaatcagcgaaatgcac	!
	\\
	N1r	
	\zz ..gtctaagttgaccgtcattggtct-5 !	Y 100\%	$T_m$=56.3$^{\circ}$C
	\zz..|||||||||||||||||||||||| !
	\zz ctcagattcaactggcagtaaccagaatgg !
	\\
	N2f
	\zz 5-ttacaaacattggccgcaaa !	Y 100\%	$T_m$=52.9$^{\circ}$C
	\zz..|||||||||||||||||||| !
	\zz gattacaaacattggccgcaaattgc !
	\\
	N2r
	\zz..aagaagccttacagcgcg-5 !	Y 97\%	$T_m$=52.2$^{\circ}$C
	\zz..|||||||||||||||||:!
	\zz cgttcttcggaatgtcgcgtattg!
	\\
	\newline
	EPI ISL 415600
	\\
	N1f
	\zz 5-gaccccaaaatcagcgaaat !	Y 100\%	$T_m$=51.6$^{\circ}$C
	\zz..|||||||||||||||||||| !
	\zz tggaccccaaaatcagcgaaatgcac !
	\\
	N1r
	\zz..gtctaagttgaccgtcattggtct-5 !		N 97\%	$T_m$=51.7$^{\circ}$C
	\zz..|||||||||:|||||||||||||| !
	\zz ctcagattcaattggcagtaaccagaatgg !
	\\
	N2f
	\zz 5-ttacaaacattggccgcaaa !	Y 100\%	$T_m$=52.9$^{\circ}$C
	\zz..|||||||||||||||||||| !
	\zz gattacaaacattggccgcaaattgc !
	\\
	N2r
	\zz..aagaagccttacagcgcg-5 !	Y 100\%	 $T_m$=54.7$^{\circ}$C
	\zz..|||||||||||||||||| !
	\zz cgttcttcggaatgtcgcgcattg !
	\\
	\newline
	
	EPI ISL 416650
	\\
	N1f
	\zz 5-gaccccaaaatcagcgaaat !	N 97\%	$T_m$=42.6$^{\circ}$C
	\zz..|||||||||||||:|||||| !
	\zz tggaccccaaaatcatcgaaatgcac !
	\\
	N1r
	\zz..gtctaagttgaccgtcattggtct-5 !		Y 100\%	$T_m$=56.3$^{\circ}$C
	\zz..|||||||||||||||||||||||| !
	\zz ctcagattcaactggcagtaaccagaatgg !
	\\
	N2f
	\zz 5-ttacaaacattggccgcaaa !	Y 100\%	$T_m$=52.9$^{\circ}$C
	\zz..|||||||||||||||||||| !
	\zz gattacaaacattggccgcaaattgc !
	\\
	N2r
	\zz..aagaagccttacagcgcg-5 !	Y 100\%	$T_m$=54.7$^{\circ}$C
	\zz..|||||||||||||||||| !
	\zz cgttcttcggaatgtcgcgcattg !
	\\
	\newline
	
	EPI ISL 417938
	\\
	N1f
	\zz 5-gaccccaaaatcagcgaaat !	Y 100\%	$T_m$=51.6$^{\circ}$C
	\zz..|||||||||||||||||||| !
	\zz tggaccccaaaatcagcgaaatgcac !
	\\
	N1r
	\zz..gtctaagttgaccgtcattggtct-5 !		Y 100\%	$T_m$=56.3$^{\circ}$C
	\zz..|||||||||||||||||||||||| !
	\zz ctcagattcaactggcagtaaccagaatgg !
	\\
	N2f
	\zz 5-ttacaaacattggccgcaaa !	Y 95\%	$T_m$=47.3$^{\circ}$C
	\zz..|||:|||||||||||||||| !
	\zz gattataaacattggccgcaaattgc !
	\\
	N2r
	\zz..aagaagccttacagcgcg-5 !	Y 100\%	$T_m$=54.7$^{\circ}$C
	\zz..|||||||||||||||||| !
	\zz cgttcttcggaatgtcgcgcattg !
	\\
	\newline
	
	EPI ISL 422983
	\\
	N1f
	\zz 5-gaccccaaaatcagcgaaat !	Y 95\%	$T_m$=44.6$^{\circ}$C
	\zz..||:||||||||||||||||| !
	\zz tggaacccaaaatcagcgaaatgcac !
	\\
	N1r
	\zz..gtctaagttgaccgtcattggtct-5 !		N 95\%	$T_m$=52.1$^{\circ}$C
	\zz..|||||:|||||||||||||||||| !
	\zz ctcagatacaactggcagtaaccagaatgg !
	\\
	N2f
	\zz 5-ttacaaacattggccgcaaa !	Y 100\%	$T_m$=52.9$^{\circ}$C
	\zz..|||||||||||||||||||| !
	\zz gattacaaacattggccgcaaattgc !
	\\
	N2r
	\zz..aagaagccttacagcgcg-5 !	Y 100\%	$T_m$=54.7$^{\circ}$C
	\zz..|||||||||||||||||| !
	\zz cgttcttcggaatgtcgcgcattg !
	\\
	\newline
	
	EPI ISL 424955
	\\
	N1f
	\zz 5-gaccccaaaatcagcgaaat !	Y 100\%	$T_m$=51.6$^{\circ}$C
	\zz..|||||||||||||||||||| !
	\zz tggaccccaaaatcagcgaaatgcac !
	\\
	N1r
	\zz..gtctaagttgaccgtcattggtct-5 !		Y 100\%	$T_m$=56.3$^{\circ}$C
	\zz..|||||||||||||||||||||||| !
	\zz ctcagattcaactggcagtaaccagaatgg !
	\\
	N2f
	\zz 5-ttacaaacattggccgcaaa !	N 97\%	$T_m$=43.1$^{\circ}$C
	\zz..|||||||||||||||:|||| !
	\zz gattacaaacattggcctcaaattgc !
	\\
	N2r
	\zz..aagaagccttacagcgcg-5 !	Y 100\%	$T_m$=54.7$^{\circ}$C
	\zz..|||||||||||||||||| !
	\zz cgttcttcggaatgtcgcgcattg !
	\\
	\newline
	
	EPI ISL425148
	\\
	n1f
	\zz 5-gaccccaaaatcagcgaaat !	Y 100\%	$T_m$=51.6$^{\circ}$C
	\zz..|||||||||||||||||||| !
	\zz tggaccccaaaatcagcgaaatgcac !
	\\
	N1r
	\zz..gtctaagttgaccgtcattggtct-5 !		N 97\%	$T_m$=51.7$^{\circ}$C
	\zz..|||||||||:|||||||||||||| !
	\zz ctcagattcaattggcagtaaccagaatgg !
	\\
	N2f
	\zz 5-ttacaaacattggccgcaaa !	Y 100\%	$T_m$=52.9$^{\circ}$C
	\zz..|||||||||||||||||||| !
	\zz gattacaaacattggccgcaaattgc !
	\\
	N2r
	\zz..aagaagccttacagcgcg-5 !	Y 100\%	$T_m$=54.7$^{\circ}$C
	\zz..|||||||||||||||||| !
	\zz cgttcttcggaatgtcgcgcattg !
	\\

	\clearpage

\section*{Acknowledgment}

The authors are grateful to Sergei Maslov and Nigel Goldenfeld from the University of Illinois for several insightful discussions.

\ifCLASSOPTIONcaptionsoff
  \newpage
\fi

\clearpage

\bibliographystyle{IEEEtran}
\bibliography{good-turing,covidgt-refs,bare_jrnl}

\newpage

\end{document}